\documentclass{article}

\usepackage{arxiv}
\usepackage{amssymb}
\usepackage{epigraph} 
\usepackage{listings}
\usepackage{longtable}
\usepackage{float}

\usepackage[utf8]{inputenc} 
\usepackage[T1]{fontenc}    
\usepackage{hyperref}       
\usepackage{url}            
\usepackage{booktabs}       
\usepackage{amsfonts}       
\usepackage{nicefrac}       
\usepackage{microtype}      
\usepackage{lipsum}
\usepackage{graphicx}
\usepackage{booktabs}
\usepackage{multirow}
\usepackage{siunitx}

\title{Promoting Rigour in Blockchain’s Energy \& Environmental Footprint Research: A Systematic Literature Review}

\author{
  Ashish Rajendra Sai \\
  Department of Computer Science\\
  Open Universiteit, Netherlands\\
  \&\\
  SCET, University of California, Berkeley\\
  \texttt{ashish.sai@ou.nl} \\
   \And
 Harald Vranken \\
  Department of Computer Science\\
  Open Universiteit, Netherlands\\
 \&\\
 Institute for Computing \& Information Sciences\\
 Radboud University, Netherlands
}

\begin{document}
\maketitle

\begin{abstract}

There is a growing interest in understanding the energy and environmental footprint of digital currencies, specifically in cryptocurrencies such as Bitcoin and Ethereum. These cryptocurrencies are operated by a geographically distributed network of computing nodes, making it hard to accurately estimate their energy consumption.
Existing studies, both in academia and industry, attempt to model the cryptocurrencies energy consumption often based on a number of assumptions for instance about the hardware in use or geographic distribution of the computing nodes. A number of these studies has already been widely criticized for their design choices and subsequent over or under-estimation of the energy use.

In this study, we evaluate the reliability of prior models and estimates by leveraging existing scientific literature from fields cognizant of blockchain such as social energy sciences and information systems. We first design a quality assessment framework based on existing research, we then conduct a systematic literature review examining scientific and non-academic literature demonstrating common issues and potential avenues of addressing these issues.

Our goal with this article is to to advance the field by promoting scientific rigor in studies focusing on Blockchain's energy footprint. To that end, we provide a novel set of codes of conduct for the five most widely used research methodologies: quantitative energy modeling, literature reviews, data analysis \& statistics, case studies, and experiments. We envision that these codes of conduct would assist in standardizing the design and assessment of studies focusing on blockchain-based systems' energy and environmental footprint.

\end{abstract}


\section{Introduction \label{1introduction}}





\epigraph{All models are wrong, but some are useful.}{George E. P. Box}
This famous quote from the British statistician George E. P. Box highlights both the merit and limits of statistical modeling. All models designed to represent real-world systems are inherently limited due to their reductive nature, however, they may serve a useful purpose if designed and tested well and if their scope and assumptions are clearly indicated. This is particularly true in the case of energy consumption models designed for sociotechnical systems \cite{orgerie2014survey}. Designing these models is a non-trivial task that requires a number of social, economic, and technical assumptions. The intent behind many of these models is often to provide a useful insight in the form of an estimate of the the energy requirement or environmental footprint, rather than an absolute measurement of energy consumed or carbon emission by these systems. 
 
Some of the early models from 2000s predicted the energy requirements of the internet and computers to a varying degree of accuracy. With some early reports suggesting that all computers could consume up to 50\% of U.S. electricity in 2010 \cite{stephens}. These claims have since been debunked through further research and empirical data \cite{koomey2008turning}. This pattern of inaccurate or misleading predictions and measurements regarding the energy consumption of a fast-growing information technology is considered problematic as it may influence policymakers \cite{koomey2002sorry} and may feed misinformation to the general public when picked up by popular media. 
  
  
  Decentralized digital assets are one such class of fast-growing information technology that have garnered significant interest from both academia and industry due to its unique energy profile \cite{sedlmeir2020energy}. Bitcoin and other similar decentralized digital assets often employ an energy-intensive consensus mechanism\footnote{In distributed computing systems such as Peer-to-Peer network-based cryptocurrencies, a consensus mechanism is employed to achieve an agreement on a single view of the data such as a ledger of transactions. We refer the reader to \cite{zheng2018blockchain}, for further information on consensus mechanisms in blockchain-based systems.} known as Proof-of-Work. 
  
  By its design, the participants in the Proof-of-Work based digital assets are incentivized to spend considerable effort, typically by executing compute-intensive or memory-intensive tasks, on a dynamically calibrated problem\footnote{In Bitcoin like Proof-of-Work based cryptocurrencies, the participants are tasked with the problem to find a block hash value below a set threshold. The difficulty of this problem is periodically changed to maintain the system property of 10 minutes time difference between two blocks of transactions.}. The first participant to find and broadcast the solution to this problem within a dedicated time frame, is rewarded for their participation in the form of newly minted cryptocurrencies. For example, on 1st June 2022, the reward to find the solution or to mine one Bitcoin block was around 200K USD (\cite{bitcoinvisuals}). This high reward induces an arms race to mine the next block by spending more computational cycles on the problem. Each attempt to find a solution to the problem incurs an energy cost in the form of electricity spent to power the device that solves the problem. 
  
  
  Similar to the early days of the internet and computers, we have seen numerous attempts at measuring the electricity consumption of decentralized digital assets such as Bitcoin \cite{lei2021best}. It has been a frequent sight to see news headlines indicating the colossal energy and environmental footprint of Bitcoin. Many of the non-academic literature and (highly rated) academic sources used in these news headlines have been criticized for inaccuracy or misleading interpretations (\cite{bevand_2017,masanet2019implausible,koomey2019estimating}). 
  
  While we acknowledge that it is worthwhile to explore the energy and environmental footprint of cryptocurrencies such as Bitcoin, we stress that this should be done with utmost care to avoid inaccurate analysis and unjustified assumptions that may lead to sensational news headlines. For instance, the article published by \cite{mora2018bitcoin} suggested that Bitcoin alone could push global warming above 2 degrees Celcius as soon as 2033. This article has been widely criticized for provably inaccurate underlying assumptions such as participants using unprofitable hardware (\cite{masanet2019implausible,koomey2019estimating,houy2019rational,dittmar2019could}). 
  
  As it is inherent with energy modeling, each of these models rely on several assumptions to provide an estimate, thus their accuracy is subject to the validity of their underlying assumptions. The scientific expectation is that these assumptions are not only mentioned explicitly but also be backed by verifiable, preferably empirical evidence or justification (\cite{sovacool2018promoting}). 


Unfortunately as seen in the case of \cite{mora2018bitcoin}, it is not always the case. Further research into the reliability of these studies by \cite{koomey2019estimating,lei2021best} has suggested that these issues are not isolated to one particular study. However as they both have only focused on a small set of models, it is difficult to generalize the results to the whole field.

Our study attempts to overcome this limitation by conducting a systematic literature review of both scientific and non-academic literature focusing on the energy and environmental footprint of cryptocurrencies.  We assess the quality of the shortlisted literature against the guidelines put forth by Lei et al. (2021) and Sovacool et al. (2018) \cite{sovacool2018promoting}. 

We iteratively refine our quality assessment framework to account for domain-specific variations \footnote{This is particularly important for the guidelines provided by Sovacool et al. (2018), as these guidelines are not specific to the blockchain domain.}. Thus, in this work, we present the first in-depth analysis of scientific rigor of blockchain energy and environmental models in order to assess the following question:

\textit{Are the existing energy and environmental footprint models and resulting estimates for blockchain-based systems trustworthy?}

It is important to note that the purpose of our article is not to discuss whether or to what extent specific studies are flawed but to provide tools to transparently discuss the rigor of these studies while allowing for improvements in the design and prediction of these models. We support and encourage the work done in Blockchain energy sciences over the last few years and intend to expand on it through this review. 

To study the reliability of energy models, we first coded and analyzed relevant scientific and non-academic literature. We review the literature published from 2008 on, i.e. post the introduction of Bitcoin's white paper \cite{nakamoto2008bitcoin}.  This is done by following the guidelines proposed by \cite{kitchenham2007guidelines}. As suggested by Kitchenham et al. (2007), our review is broken down into five steps: \textit{Search, Selection, Quality Assessment, Data Extraction, and Analysis}. This review results in an article pool of 128 studies. These articles are then assessed for their scientific rigor by using the quality assessment framework based on the guidelines of \cite{sovacool2018promoting} and \cite{lei2021best}. 

Following the assessment of the scientific rigor, we consolidate our findings in the form of commonly occurring issues in different types of studies. We also document potential avenues of addressing these known issues. This subsequently leads to the development of novel code of practices to promote scientific rigor in blockchain energy studies. 

We believe that this study assists the reader in understanding the reliability of the current energy and environmental studies in a blockchain context. This article also assists researchers and developers in designing or refining their existing models through adherence to the code of practice. The paper makes the following contributions:

\begin{itemize}
    \item We systematically review the existing literature to document common issues with energy and environmental impact studies for Blockchain-based systems (Section \ref{3methodology}).
    \item We develop a novel quality assessment framework for Blockchain-specific studies that can assist in understanding the scientific rigor of the energy or environmental impact model (Section \ref{3methodology}). 
    \item We identify research gaps specifically with regards to the lack of non-Bitcoin-specific investigations in academic literature. We also report on the lack of empirical data for these models (Section \ref{4quality}). 
    \item We manifest the findings of our review in a set of best practices that can assist in designing or improving existing models (Section \ref{COP}). \ref{discussion}). 

\end{itemize}

\section{Background \label{2background}}
Cryptocurrencies that use an energy-intensive consensus mechanism cause two prime concerns from an environmental perspective: the electricity consumption and the carbon emission associated with the energy consumption\footnote{There are other environmental impacts associated with cryptocurrency operations such as E-Waste generation \cite{de2021bitcoin}, we briefly touch on this in a subsequent section however our focus in this article is primarily on energy consumption and carbon emissions}. In this section, we provide an overview of how the energy and environmental footprints of these cryptocurrencies are usually measured. 

\subsection{Energy Consumption}
As alluded to in the introduction section, measuring the energy consumption of a geographically distributed network is a non-trivial task. This problem is compounded when considering decentralized systems as it is difficult to find a centralized source of information about the network's physical composition \cite{sai2021taxonomy}. There are two main ways of estimating the energy consumption of a blockchain-based system depending on the availability of reliable data on the computing network: \textit{bottom-up} and \textit{top-down}. 

\subsubsection{Bottom-Up}
A distributed computing network is made up of computational devices that consume a certain amount of electricity per unit of work\footnote{In Proof-of-Work based cryptocurrencies, the work is often performing hashing operations to find a nonce value such that the resulting hash value is lower than the target. }. Each of these computational devices can have different performance and energy efficiency profiles. For example, a network could be made up of 100 Raspberry Pi\footnote{See www.raspberrypi.com.} devices generating X unit of work in a single unit of time or it could be made up of 2 consumer-level personal computers generating the same amount of work in the same time resolution. 

One of the early attempts at using a bottom-up approach for modeling electricity consumption of Bitcoin was made by \cite{bevand}. In his analysis, \cite{bevand} outlined prominent modern bitcoin mining hardware that typically employ application-specific integrated circuits (ASICs) designed for bitcoin mining. For example, an Antminer S9 system released in 2017 could perform 13 TH/s whereas a consumer CPU such as Intel i7 (2021) can only perform 2.5 KH/s while being more power efficient per unit of hash calculation. 

If we are aware of the exact hardware used in the network, including the hardware distribution (how many units of each type of device are on the network), we can use this information to calculate the total energy consumed by all the constituting computing devices. 

This calculation requires accurate values of each device's computing power and energy efficiency. This in itself can be problematic in a real-world scenario, as most of the information about power and energy efficiency is obtained through data sheets provided by manufacturers. These data sheets in most cases are not verified by an independent auditor. Furthermore, tuning operational parameters like clock frequency and supply voltage may even lead to different numbers in practice. For an accurate measurement, it is also important to know the uptime for each device and the actual work done during this uptime. 

This gives us a partial understanding of the network's energy consumption as this calculation does not consider operational electricity consumption for devices other than the computing device such as the networking or cooling infrastructure. These operational expenses are often considered in the form of a fractional value known as Power Usage Effectiveness (PUE) \cite{brady2013case}.

Once we know the energy consumption of each device in the network and the associated PUE value, we can calculate the total energy consumed as follows \footnote{It is worth noting that this equation is for illustrative purpose only as in the real model, the authors might account for additional factors such as economic of operators.}:

\begin{equation}
    T = \sum {\varepsilon(i)}* PUE(i)
\end{equation}

\begin{figure}
    \centering
    \includegraphics[scale=0.40]{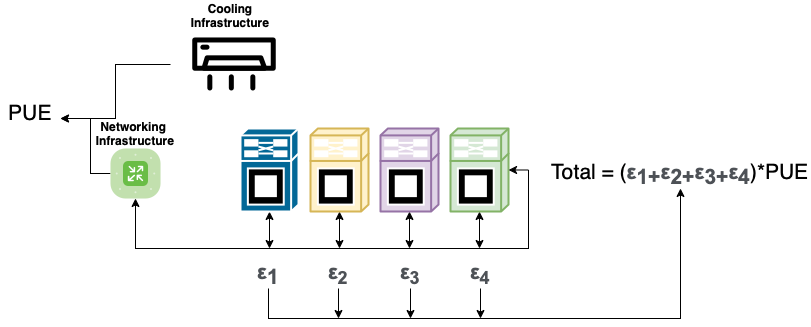}
    \caption{Calculating Energy Consumption using Bottom Up Approach}
    \label{fig:BU}
\end{figure}

Where $T$ is the total energy consumption, $\varepsilon$ is the energy consumption of each constituting computing device and PUE is the additional operational electricity requirement. This calculation is also visually illustrated in Figure \ref{fig:BU}.

\subsubsection{Top-Down}
Bitcoin and other cryptocurrencies are often described as decentralized systems. Decentralization is a crucial component for the network on different levels of operations such as applications (decentralized exchanges), protocol (consensus mechanism), and network (distributed peer-to-peer network) \cite{sai2021taxonomy}. This decentralized nature of the network makes it difficult for researchers to collect empirical data on the location and hardware of the consensus participants\footnote{In Proof-of-Work based systems, these consensus participants are also known as miners.}. 

Due to the lack of reliable empirical information on the network participants, a large number of energy and environmental models are based on a top-down approach \cite{ccaf}. In a top-down modeling approach, the model relies on high-level technical, economical or social variables. 

For instance, \cite{vranken2017sustainability}'s top-down model is based on the total computing power also known as hash rate. In \cite{de2018bitcoin}'s model, the author builds an economic model based on the economic rationality of the miner. 

In this subsection, we provide an abstract overview of how a Top-Down model conceptually works, however, we refer the reader to \cite{lei2021best,ccaf,vranken2017sustainability} for an in-depth discussion of top-down modeling. 

For Bitcoin, we can calculate the total hash rate of the network by using the difficulty of mining \cite{o2014bitcoin}. First of all, an estimate on the hash rate of the network is required. This can be derived from a simple statistical model that considers the difficulty parameter and the time it takes on average to mine a block, which is 10 minutes for Bitcoin, as is applied by \cite{o2014bitcoin}. A more refined model may use empirical data on the exact amount of time it takes to mine blocks. For instance, for Bitcoin the difficulty parameter is adjusted every 2,016 blocks, and hence some drift may occur in between.  

The total hash rate of the network is composed by the combined hash rates of a number of different hardware in use, each with a specific energy and performance profile. A number of different combinations and permutations of available hardware can generate the required hash rate. Different models make different assumptions in order to get to the total hash rate. For example, some models assume that the network is made up of only the most efficient commercially available hardware while others consider a pool of hardware with different distributions. We can represent this calculation as follows \footnote{There are often many different combinations of devices possible here with different $\rho$. For instance, a small network can be made up of a large number of inefficient devices or a small number of highly efficient devices.}:

\begin{equation}
    \textbf{H} = \sum \rho 
\end{equation}

Here $H$ is the hashing power of the network composed of all the individual hashing power ($\rho$) of the hardware used in the network. 
Once a pool of hardware is decided upon, we can similarly calculate the energy consumption to that of bottom-up\footnote{The $\varepsilon$ is only for hardware that contribute to $H$ above.}:
\begin{equation}
     T = \sum {\varepsilon(i)}*PUE(i)
\end{equation}

\begin{figure}
    \centering
    \includegraphics[scale=0.4]{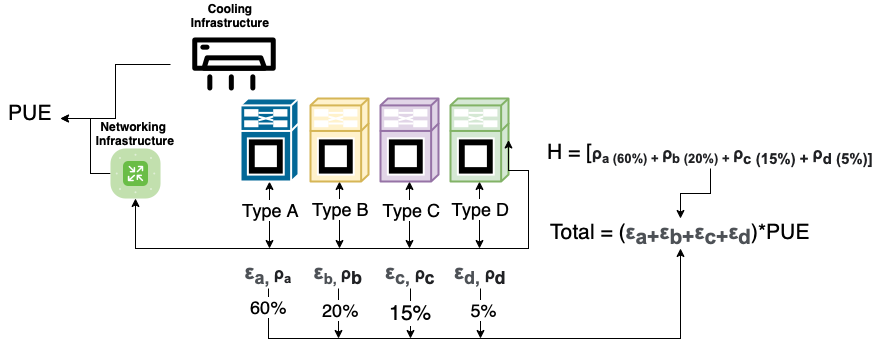}
    \caption{Calculating Energy Consumption using a Top Down approach}
    \label{fig:TD}
\end{figure}

We have also illustrated this process in Figure \ref{fig:TD}.

\subsection{Environmental Impact Measurement}
The scope of the environmental impact of information technologies can be very broad ranging from the direct impact caused by E-waste and through the consumption of electricity generated by non-renewable carbon intensive operations such as coal-based power plants \cite{k2019}. Through our literature review, we report that most of the studies in the blockchain context focus on the carbon emission associated with the electricity consumption of the network. 
However, it is worth noting that there are a few studies that look at other aspects of the environmental impact of cryptocurrencies such as E-waste generation \cite{de2021bitcoin} and scope 2 and 3 carbon emissions \cite{johnson_pingali}. In this subsection, we focus on carbon emissions.

The carbon emission calculation consists of a five-step process as outlined below\footnote{It is important to note that these steps are only indicative of the process, individual studies may differ in their approach.}:

\begin{enumerate}
		\item Calculating the total energy consumed by the network: This can be determined by either Bottom-Up or Top-Down approaches as discussed above.
		\item Determining the geographic location of the devices in the network: In addition to understanding the pool of hardware and their respective share of the total network, we also need to know their geographic location. 
		\item Obtaining the energy mix for shortlisted geographic locations: For each geographic location, we need to understand the breakdown of the electricity in the form of its sources of energy. An electricity grid in a country with heavy reliance on renewable energy is likely to produce considerably less carbon per unit of electricity than a country with a coal-based electricity grid. 
		\item Use grid emission factor to calculate the $CO_2$ emission: A geographic area with a predominately green grid should account for less carbon emission per unit of electricity generation and consumption, while one with reliance on coal power should have high carbon emission. This is captured in the from of a grid emission factor. The grid emission factor measures the amount of $CO2$ emissions per unit of electricity in a given geographic area. 
		\item Sum all the carbon emissions: In this step, we generate a final carbon emission value for the whole network by adding all the individual carbon emission data points from each constituting geographic location. 
	\end{enumerate}

\subsection{Sensitivity of these models}
In the preceding subsections, we provided a brief overview of the prominent methodologies used to calculate both the energy and environmental impact of cryptocurrencies. It is important to note that the Bottom-Up method despite seeming straightforward is quite difficult to execute for large cryptocurrencies due to the lack of reliable data. This may be due to the inherent pseudo-anonymous structure of these cryptocurrencies and the reluctance of participants to disclose their identity due to the fear of governmental retaliation \cite{huckle2016socialism}. 

Thus most of the current studies are based on the Top-Down modeling approach which requires a number of assumptions that may impact the final estimate of the model. For example, one of the most widely used models for Bitcoin is designed by Cambridge Centre for Alternative Finance \cite{ccaf}. This model attempts to use the profitability of mining to calculate the total power consumption of the network. In doing so, the authors make a number of assumptions including an assumption that on average the miners pay 0.05 USD/KWh. If we change the cost of electricity to 0.10 USD/KWh\footnote{This change is in line with other studies focusing on Bitcoin's energy consumption. We will discuss how the cost of electricity varies significantly in different studies.} the final model estimates drops by about 38\% \footnote{This calculation is based on data collected on July 23, 2022.}. This is a significant difference based on the change in a single variable, this alludes to the sensitivity of the model on its parameters. 

Unlike the CCAF model, where most of the parameters are based on empirical assumptions other models do not fair well. For instance, the models developed by Mora et al. (2018) and De Vries (2018) have been widely criticized for their underlying assumptions \cite{masanet2019implausible,koomey2019estimating,houy2019rational}. 

In this article, we attempt to systematically document these variables and their impact on accuracy to allow researchers a means to improve their models or representation of the results from these models. 

\section{Methodology\label{3methodology}}
In this section, we provide an overview of the methodology used to conduct the systematic literature review, quality assessment framework and code of practices.

Due to the diverse nature of blockchain research, it is difficult to come up with a single ``ten steps to quality" measure. Thus, we begin our search by identifying and classifying prominent research methodologies employed in blockchain energy sciences. To this end, we define two primary research questions for our literature search:

\begin{enumerate}
    \item \textbf{RQ1}: What are the different methods used to measure/model the energy or environmental footprint of blockchain-based systems and their associated modeling assumptions?
    \item \textbf{RQ2}: What are the implicitly or explicitly acknowledged strengths or limitations of these models? 
\end{enumerate}

We attempt to answer these questions through a systematic literature review. For the systematic literature review, we follow the guidelines proposed by \cite{kitchenham2007guidelines} to identify relevant literature in both academic and non-academic domains. Based on the literature review, we iteratively refine and generate a novel quality assessment framework based on the guidelines proposed put forth by \cite{sovacool2018promoting} and \cite{lei2021best}. 

\begin{figure}
    \centering
    \includegraphics[scale=0.4]{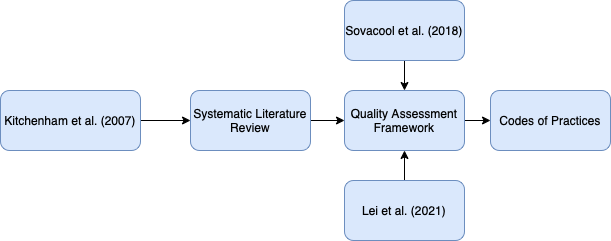}
    \caption{Research Flow}
    \label{fig:my_label}
\end{figure}

We then apply this quality assessment framework to the shortlisted studies and document common issues with the shortlisted studies. These issues form the basis for the codes of practices proposed later in the study. These code of practices are based on well-established standards from information systems, statistics, social energy sciences, and blockchain-specific literature. The flow of this study is illustrated in Figure \ref{fig:my_label}.

\subsection{Systematic Literature Review}
As indicated earlier, we follow the guidelines put forth by \cite{kitchenham2007guidelines}. We conduct our review in 3 phases: in the first phase, we construct the search query by explicitly documenting search strings. 

\begin{figure}
    \centering
    \includegraphics[scale=0.29]{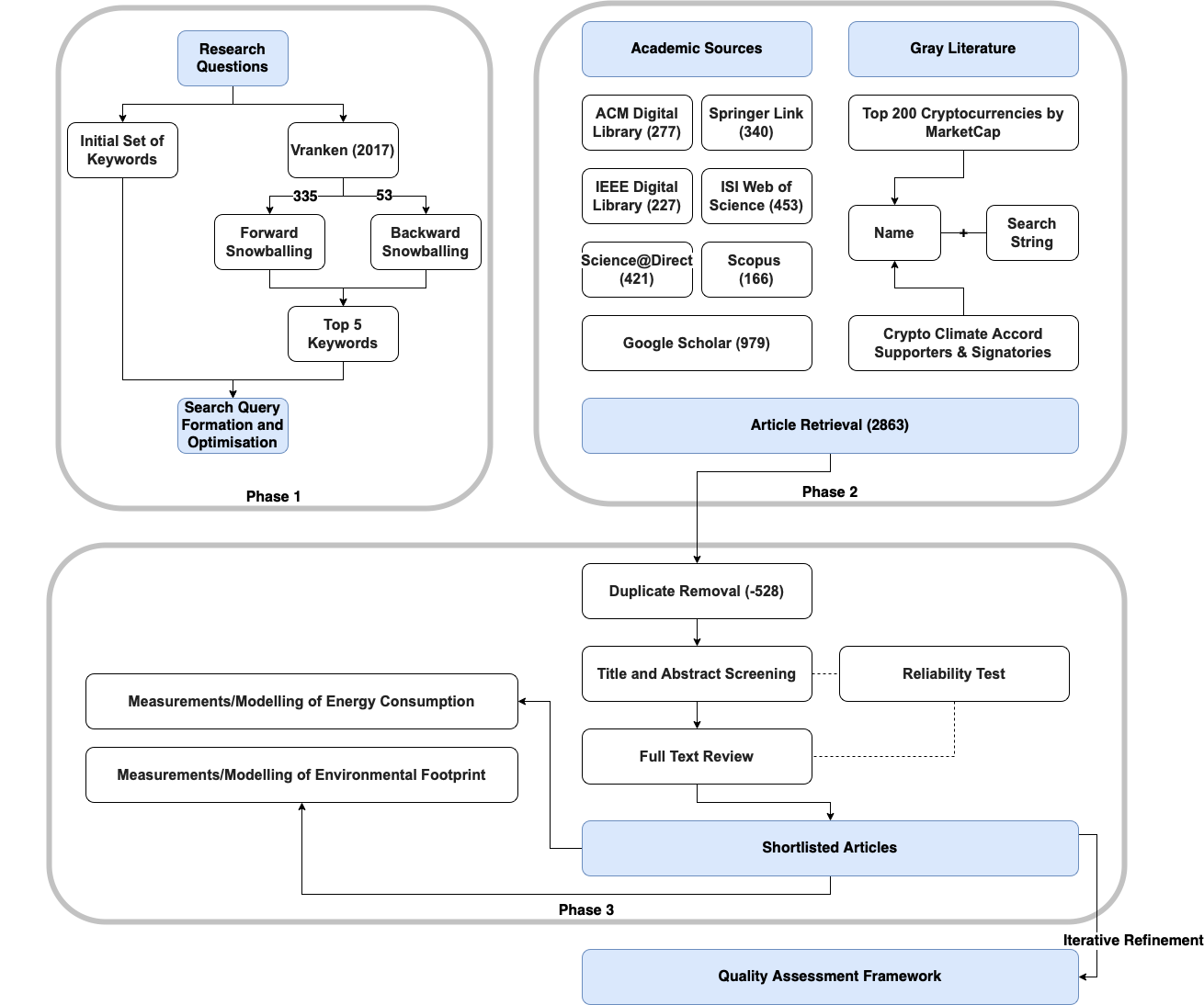}
    \caption{Systematic Literature Review Protocol}
    \label{fig:SLR}
\end{figure}

In the second phase, we conduct the search for relevant articles by first shortlisting appropriate repositories and sources of academic and non academic literature. In the final phase we extract the measurement/modeling technique used for energy/environmental footprint analysis from the shortlisted articles. These three phases are illustrated in Figure \ref{fig:SLR}.

\subsubsection{Phase 1: Search Query Formation}

RQ1 increases the coverage of our research by capturing different models and their associated assumptions. If a shortlisted article proposes a new model with assumptions, we include these new variables in our quality assessment framework. For each identified variable, we document all the assumptions made by the study. 

With RQ2, we further solidify our understanding of the limitations associated with these models by documenting the acknowledged strengths and weaknesses of each shortlisted model. This documentation helps us prepare our code of practice. 

These two primary research questions serve as the basis of our literature search strategy. We start by constructing a search string to identify relevant literature. We do this by first constructing an initial set of keywords based on both the RQs and then further validating these keywords by performing backward and forward snowballing on \cite{vranken2017sustainability}\footnote{We have plotted a word cloud from the backward and forward snowballing in \ref{AppendixA}.}.  We choose \cite{vranken2017sustainability} for snowballing as it is one of the first papers in the field while being one of the highest cited. The final search query is of the following structure:

\begin{lstlisting}
Blockchain: "Blockchain" OR "DLT" OR "bitcoin" OR "blockchain" OR "cryptocurrencies" OR "cryptocurrency" OR "digital currency" OR "distributed ledger" OR "peer-to-peer computing" OR "smart contract platform"
Energy: "electricity" OR "power" OR "power supply"
Consumption: "expenditure" OR "use" OR "utilisation" OR "utilization"
Environment: "atmosphere" OR "carbon" OR "climate" OR "ecological" OR "emission" OR "environmental" OR "green" OR "footprint" OR "e-waste"
Sustainability: "green design" OR "green technology" OR "sustainable"
\end{lstlisting}

\subsubsection{Phase 2: Article Search}

In article search, we intend to ensure high coverage of models and their associated limitations and strengths, we do this by extracting relevant models from both academic and non-academic literature sources. For academic sources, we search prominent computer and energy science research repositories: \textbf{Google Scholar, ACM Digital Library, IEEE Digital Library, ISI Web of Science, Science Direct, Scopus, and Springer Link}\footnote{The exact search query used for each of these platforms is reported in \ref{AppendixB}.}. Out of these search repositories, Google scholar returned the highest number of articles due to its wide scope and coverage of potential non-academic literature as well. However, due to the limitations imposed by Google, it is only feasible to obtain the first 980 results. It is worth noting that the relevance of results depletes significantly in Google scholar after the initial few articles \cite{sai2021taxonomy}. 

For non-academic literature results, we identify two primary sources of information: the cryptocurrencies consuming the energy and third parties providing services around these cryptocurrencies. We conduct an exhaustive search of the top 200 cryptocurrencies by market cap using the Google search engine to locate any energy or environmental footprint models developed for these cryptocurrencies\footnote{A complete list of cryptocurrenices reviewed is present in \ref{AppendixC}}. 

Due to the vast number of results, we limit our Google search to only the first 100 results. For capturing the models developed by service providers, we consult the list of supporters and signatories of the Crypto Climate Accord\footnote{Crypto Climate Accord is a recent industry-driven accord to limit the environmental impact of blockchain technologies. As of 01-June-2022, there are 200 supports and 150 signatories to the accord.}. 

As the nature of our query is left intentionally vague to improve coverage, we end up with an initial set of 2863 articles. Out of these 2873, a majority (2614) of the results are from academic sources. The non-academic literature consists of 259 results mostly in the forms of blog posts and whitepapers \footnote{The actual number of non-academic literature included in our study is likely significantly high due to the inclusion of results from Google Scholar which tends to capture some non-academic literature sources such as academic blogs.}. In the next phase, we document the process used to shortlist the relevant articles. 
\subsubsection{Phase 3: Shortlisting Relevant Articles}

Due to a large number of articles returned for our initial search, we conduct a title and abstract screening to shortlist the articles using explicit inclusion criteria as outlined below: 
\begin{enumerate}
    \item The paper’s title mentions energy or environment, or any of the synonyms mentioned in 3.1.1, or is potentially relevant to the study of blockchains energy or environmental impact research.
    \item The abstract is relevant to the measurement of energy or environmental footprint.
\end{enumerate}

This filteration was done by the first author and resulted in 244 potentially relevant articles. To ensure that the title and abstract review process is reliable, we conduct a reliability test. To do this, we performed cross-validation using the mechanism proposed by \cite{fleiss1973equivalence}. We perform the cross-validation on a sample of 67 with a confidence interval of \textbf{90\%} and margin of error of \textbf{10\%} calculated in line with the suggestions of  \cite{sim2005kappa}. These shortlisted 67 articles were screened and reviewed by the second author independently to perform cross-validation. The results from the cross-validation indicate a substantial agreement between the authors with Cohen's Kappa value of 0.71. This suggests that the shortlisting process is reliable. 

The title and abstraction process significantly reduces the number of articles in our search pool to a set of 244 articles. We then review the full text of these 244 articles to assess the relevance. In order to assess their relevance to our survey, we define a relevance matrix for shortlisting in Table \ref{tab:relavanceM}. A study that complies with any one of these three factors is included in our final set of relevant articles. This results in the final set of 128 studies\footnote{A complete list of all the shortlisted studies is included in \ref{AppendixD} } that are considered directly relevant to our research.

\begin{table}[]
    \centering
    \begin{tabular}{|p{8cm}|l|l|}
    \hline
        Factor & Yes & No \\ \hline
         Measurement or modeling technique identified for energy consumption & 0.5 & 0 \\ \hline
         Measurement or modeling technique identified for environmental footprint & 0.5 & 0 \\ \hline
         Acknowledgment of strength and/or weakness of an energy/environmental footprint model & 0.5 & 0\\ \hline
         
    \end{tabular}
    \caption{Relevance Matrix}
    \label{tab:relavanceM}
\end{table}

\subsection{Development of Quality Assessment Framework}

Objectively assessing and discussing the scientific rigor of a field as diverse as blockchain is a difficult exercise mostly due to the diverse nature of the research and application of this technology. In this study, we attempt to understand the common issues in the design and execution of these studies and provide guidelines on avoiding these mistakes. To do this, we first adopt the basic research elements proposed by \cite{sovacool2018promoting} into four quality indicators: \textit{clear research question, building upon existing knowledge, explicit research design, and reliability assessment of the underlying data}. We document these four basic research quality indicators and their associated measurement techniques in Table \ref{tab:BRQI}.

These four indicators of basic research quality allow us to discuss research design independent of underlying research methodologies. Out of these four indicators, clear research questions and building upon existing knowledge are more applicable to scientific literature while the last two are equally applicable to studies in the non-academic literature as well. 

It is expected that a scientific study should contain a research question, ideally explicitly stated within the text body \cite{kothari2004research}. The second indicator focuses on the use of existing knowledge to inform the construction of newer models, this can be in the form of building upon existing energy or environmental footprint models within the blockchain domain or application of theories from fields such as economics or social sciences. The use of existing theory promotes the evolution of the research within a field while providing a more robust mechanism to discuss strengths and limitations. 

Another important research aspect of studies that focus on the environmental impact of technology is the reproducibility of the analysis. We capture this through the third basic research quality indicator, explicit research design. Within this indicator, we expect that a study should contain sufficient details of the study design and execution to permit independent reproduction. It is also expected that a study reliant on a non-public dataset should also share the dataset and appropriate source code where possible. 

If the study is utilizing data sources from non-peer-reviewed studies or investigations, it should include a detailed description of potential reliability issues in the data. This is particularly important in the blockchain context as the non-academic literature often contains unvetted datasets of which the reliability is not validated. 

\begin{table}[]
    \centering
    \begin{tabular}{|p{4cm}|p{4.5cm}|p{4cm}|}
\hline Quality Indicators & Measurement & Application Area \\\hline 
    \textbf{BR1}: Clear Research Question & Explicit or implicit research question & Academic Literature \\ \hline
    \textbf{BR2}: Building on existing knowledge & Use of a theory or framework from existing studies & Academic Literature \\ \hline
    \textbf{BR3}: Explicit Research Design & Clear research methodology, public data and source code (where appropriate) & Academic and Non-Academic Literature \\ \hline
    \textbf{BR4}: Reliability of External Data & Reliability assessment & Academic and Non-Academic Literature \\ \hline

    \end{tabular}
    \caption{Basic Research Quality Indicators}
    \label{tab:BRQI}
\end{table}

\subsubsection{Different Research Methodologies}
Through the full-text review of the final set of 128 studies, we identify prominent research methodologies employed in the shortlisted studies. In total, we identify five distinct research methodologies: \textit{Quantitative Energy Modeling, Literature Review, Data Analysis and Statistics, Case Studies, and Experiments}. 

Out of these five, the quantitative energy modeling approach is highly dependent on the structure of cryptocurrency and the modeling approach adopted (top-down or bottom-up). Therefore the quality indicators for quantitative energy modeling are closely tied to the variables discussed in Section 2. Unlike quantitative energy modeling, other research methodologies are not as dependent on the structure of the cryptocurrency and can benefit from more generic field-specific quality indicators. For instance, the way a literature review is conducted is independent of the subject under review. Thus we reason that it is more appropriate to employ quality indicators that focus on the method irrespective of the subject. This focus on quality indicators centered on the method allows us to examine studies that might not be specific to popular crypto-assets such as Bitcoin and Ethereum. 

In the following subsection, we briefly describe each of these research methodologies and provide an outline of the quality assessment mechanism. 

\begin{itemize}
    \item Quantitative Energy Modeling: Quantitative models rely on a robust dataset that is used in conjuncture with a social, economical, or technical model of the blockchain. For instance, the CCAF model is an example of a computational economic model while the Digiconomist model is an example of a socio-economic model. Both of these models are based on the use of quantitative analysis for the estimation of energy consumption. The exact model composition within quantitative modeling can be based on any combination of the social, economic and technical variables above thus the exact quality indicators vary significantly depending on the model formulation and the intended use. These categories are outlined in Table \ref{tab:QIQEM}. 
    
    \begin{table}[]
    \centering
  \begin{tabular}{|p{4cm}|p{6cm}|p{4cm}|}
\hline Quality Indicators & Description & Measurement \\\hline 
    \textbf{QM1}: Hardware Composition & Hardware or pool of hardware used for calculation of energy consumption& Examination of the exact hardware distribution, checking assumption related to hardware efficiency \\ \hline
    \textbf{QM2}: Geographic Data & Geographic data such as the location of miniers in the network is crucial for calculating CO2 emissions &Assessment of the methodology used for the extraction of geographic distribution \\ \hline
    \textbf{QM3}: Economics & Participating in a cryptocurrency incurs both capital expenditure to acquire hardware and related operational expenses & Analyzing the cost of electricity, PUE value, Hardware Lifespan\\ \hline
    \textbf{QM4}: Social & Incentive engineering behind the design of PoW requires the miners to be rational, this also needs to be accounted for when calculating energy consumption &Rationality of agents\\ \hline
    \textbf{QM5}: Carbon Emission Data & The carbon emission data is crucial for calculating environmental impact of these crypto-assets & Source of the data and geographical resolution  \\ \hline
    \textbf{QM6}: Time Resolution & The data used in quantitative energy modeling tends to evolve over time along with the model itself thus it is important to provide time resolution for the data along with appropriate archiving of legacy data or model parameters & Review of the data and model parameters \\ \hline
    \end{tabular}
    \caption{Quality Indicators for Quantitative Energy Modeling}
    \label{tab:QIQEM}
\end{table}
    \item Literature Review: Narrative review of the literature can provide useful insights into existing models and the potential strengths and weaknesses of these models. These reviews can be targeted toward different crowds, some of these focus exclusively on experts in the domain \cite{lei2021best} while others are a more general introduction to the energy and environmental footprint of Blockchain \cite{vranken2017sustainability}. 
    
    Literature reviews are widely used in a number of cognizant fields such as Information Systems and Computer Science thus there is a wealth of guidelines on the quality of these reviews. We shortlist a subset of these guidelines and construct our quality indicators as outlined in Table \ref{tab:QILR}.
    
    \begin{table}[]
    \centering
  \begin{tabular}{|p{5.5cm}|p{5cm}|c|}
\hline Quality Indicators & Description & Measurement \\\hline 
    \textbf{LR1}: Type of review method & Some review methods such as narrative review are considered less rigorous than a meta-analysis of the literature, we follow the guideline to classify rigor based on method type  & Review of method\\ \hline
    \textbf{LR2}: Explicit Criteria & To avoid bias, it is important to document explicitly the RQs, search strings, inclusion and exclusion criteria & Review of method \\ \hline
    \textbf{LR3}: Appropriate Search Database & In fields like Blockchain, a significant portion of the research is conducted within non-academic literature thus it is important to have a high coverage & Review of method\\ \hline
    \textbf{LR4}: Sampling Process Documentation & If only a specific sample is analyzed, it is important to document the process & Review of method\\ \hline
\end{tabular}
    \caption{Quality Indicators for Literature Review}
    \label{tab:QILR}
\end{table}
\item Data Analysis and Statistics: Unlike quantitative energy modeling, a method that relies on data analysis and basic statistics is far less sophisticated. These models often rely on univariate or multivariate analysis of variables associated with the energy or environmental footprint of cryptocurrencies. Due to the less complicated nature of these studies, we apply guidelines specific to statistics and data analysis as outlined in Table \ref{tab:QIDA}:

    \begin{table}[]
    \centering
  \begin{tabular}{|p{5.5cm}|p{5cm}|c|}
\hline Quality Indicators & Description & Measurement \\\hline 
    \textbf{DA1}: Type of analysis performed & Some analysis methods such as univariate are considerably less rigorous than a longitudinal multivariate analysis  & Review of method\\ \hline
    \textbf{DA2}: Clear Hypothesis & It is important that in statistical analysis, the hypothesis is clearly stated and evaluated & Review of method \\ \hline
    \textbf{DA3}: Practical vs statistical significance & High statistical significance does not necessarily mean that the relationship analyzed or predicted is of practical significance thus the practical significate must be discussed when presenting a statistical significance value & Review of method\\ \hline
\end{tabular}
    \caption{Quality Indicators for Data Analysis and Statistical Models}
    \label{tab:QIDA}
\end{table}
\item Case Studies: Case studies allow us to take an in-depth look at a specific instance of a broader phenomenon \cite{tsang2014case}. In energy studies of Blockchain, these case studies are often specific to cryptocurrencies and their energy or environmental footprint. There is a wealth of literature on the selection and execution of case study analysis to ensure that the studies are of high relevance to the broader field \cite{tsang2014case,benbasat1987case}. We adopt and document these guidelines in Table \ref{tab:QICS}:

 \begin{table}[]
    \centering
  \begin{tabular}{|p{5.5cm}|p{5cm}|p{4cm}|}
\hline Quality Indicators & Description & Measurement \\\hline 
    \textbf{CS1}: Case Selection & Selecting an appropriate case study is important to ensure that the results are generalizable or of extreme cases that may explain a specific phenomena  & Review of selection mechanism\\ \hline
    \textbf{CS2}: Clear Boundaries & Case studies can be very broad or hyper-specific thus it is important to document the bounds of the study & Review of method \\ \hline
    \textbf{CS3}: Measurable dependent and independent variables & If the case study is analyzing a phenomenon, it is crucial to clearly define the dependent and independent variables & Review of method\\ \hline
\end{tabular}
    \caption{Quality Indicators for Case Studies}
    \label{tab:QICS}
\end{table}

\item Experiments: In some instances where it might be difficult to extract information from observation, it may be useful to conduct experiments \cite{sovacool2018promoting}. In the Blockchain energy domain, these experiments are often done to calculate the energy consumption of a specific device and then use that information to generate measurements for a subset or the whole of the network. The quality indicators for experiments are manifested in Table \ref{tab:QIEX}.

 \begin{table}[]
    \centering
  \begin{tabular}{|p{4cm}|p{5cm}|p{4cm}|}
\hline Quality Indicators & Description & Measurement \\\hline 
    \textbf{EX1}: Representative Sample & Selection of the experiment object is crucial for the generalizability of results  & Review of selection mechanism\\ \hline
    \textbf{EX2}: Choice of setting & The settings of the experiment should closely resemble that of the real world object to ensure reliable results & Review of method \\ \hline
\end{tabular}
    \caption{Quality Indicators for Experiments}
    \label{tab:QIEX}
\end{table}

\end{itemize}

In the next section, we employ these quality indicators and assess the quality of the academic and non-academic literature shortlisted through the systematic literature review. 




\section{Results}
In this section, we provide an overview of our quality analysis. We begin by describing the trends in the literature followed by a description of the prominent results.

The field of blockchain energy science is quite new with the first academic study published in 2014. The field has only seen traction in the last 4 years. The non-academic literature has specifically seen large growth since 2020, a period also referred to as DeFi summer \cite{xu2022short}. 

The research methodologies in use have also evoloved. In the early years, most of the models proposed were quite simple data analysis and statistical models but since then these have evolved into more mature quantitative models incorporating economic, social, and technical modeling. The field has also seen evolution in each of these research methods as well. For example, we note a trend towards more sophisticated literature reviews such as meta-review. 

The biggest development in terms of influential literature has been the introduction of the Digiconomist index \cite{digiconomist_2022} and then the Cambridge Bitcoin Electricity Consumption index \cite{ccaf}. Both of these indexes now underpin several of the assumptions made by newer studies. For instance, the value of electricity cost has seen a significant change since the introduction of the CCAF index, a majority of newer studies assume the cost of electricity to be 0.05 US Cents/KWh in line with the assumption made by the Cambridge Index. 

These trends however do not indicate that the field is getting more rigorous over time. In the following subsection, we discuss the basic research quality indicators followed by more specific quality checks for the five research methodologies outlined in Section \ref{3methodology}. 

\subsection{Quality Assessment\label{4quality}}
\subsubsection{Basic Research Quality Indicators}
\begin{table}[]
    \centering
    \begin{tabular}{|c|p{7cm}|}
\hline Quality Indicators & Results \\\hline 
    \textbf{BR1}: Clear Research Question & 5\% (7) studies do not contain a research question\\ \hline
    \textbf{BR2}: Building on existing knowledge & 74\% (95) of the analyzed studies do not build upon existing theories in blockchain or any other related domain\\ \hline
    \textbf{BR3}: Explicit Research Design & 34\% (44) of the studies do not have an explicit research design, while 43\% (53) of studies do not share data whereas 67\% (85) of studies do not share source code\\ \hline
    \textbf{BR4}: Reliability of External Data & 79\% (101) studies do not discuss the reliability of external data used in their analysis\\ \hline

    \end{tabular}
    \caption{Basic Research Quality Indicator Results}
    \label{tab:BRQIR}
\end{table}

We have documented the results from our analysis in Table \ref{tab:BRQIR}. It can be seen that a handful of the shortlisted studies do not describe their research goal clearly making the document difficult to understand and contextualize. 

A bigger and more systematic issue with research in this domain is the lack of building upon existing knowledge. As it can be seen from our analysis, 74\% of the studies do not build upon existing theories. For instance, in their analysis of the life cycle of Bitcoin mining, the authors \cite{k2019}, utilized the well-established Life Cycle Assessment methodology allowing us to use existing theory to assess their estimates and independently verify their results. 

In contrast to this, the analysis conducted by studies such as \cite{de2018bitcoin} does not build on any existing theory and present their own methods without contextualizing it in existing literature in blockchain or other scientific discipline making it difficult to compare the robustness of the underlying research method. We believe that there needs to be a more coordinated approach to build upon existing knowledge within blockchain research domain as well as other cognizant research domains such as energy sciences, economics and information systems. 

Another important aspect of a reliable scientific study is the reproducibility of its results however this is not trivial in blockchain energy sciences as 34\% of the analyzed studies do not contain enough information to reproduce their analysis. This issue is even more prevalent when a supporting dataset is needed for reproduction. A large portion of studies (43\%) do not share their underlying dataset either requiring the reviewer to seek the dataset or in some cases deferring the publication of the dataset to an unspecified period in the future\footnote{The widely cited CoinShare report \cite{kimmell2022} is an example of this.}. 

Some of the analyzed studies employ mathematical models that can assist in generating a system-wide figure for energy consumption or environmental footprint. These mathematical models are often deployed in the form of an excel sheet or a computing script. In either of these situations, it is desirable to have access to the underlying calculation however a vast majority (79\%) of the studies do not provide access to their source code. 

As alluded to in the previous section, the field has seen significant interest from both academic and non-academic sources in the past 4 years resulting in a plethora of models based on different assumptions and datasets. However, the reliability of these models and the dataset is not well established in most cases. For instance, a common assumption about the lifecycle of a typical Bitcoin mining hardware is that the hardware will only be profitable for 1 to 2 years.

This assumption is based on a study by \cite{de2018bitcoin} published in Joule as a commentary. However, since its publication, this assumption has been widely criticized for its oversimplified view of mining operations and dependence on anecdotal examples to back the assumption \cite{koomey2019estimating}. 

Several studies have also established that the mining hardware used in \cite{de2018bitcoin} for example was profitable for over 4 years since its production undermining the assumptions in the initial study. However, as \cite{de2018bitcoin} work is the first to give an estimate on the lifecycle of a typical Bitcoin miner, it is still being used in numerous studies despite the potential flaws. This is one of several instances where a flawed or outdated study has been used widely to develop newer models upon. Based on our analysis, we report that a majority of the studies (79\%) analyzed rely on an external dataset but do not acknowledge any potential validity issues in these studies. 

The basic research quality indicators suggest that the field is likely evolving towards more mature research methodologies however the reproducibility of this research is hampered by poor data and source code sharing practices. It can also be seen from this analysis that most of this evaluation is done in small silos rather than the whole field evolving together. This is most likely due to the lack of re-use of theory and high dependence on some non-academic or unvalidated datasets. We address these potential research limitations in Section \ref{conclusion}, where we propose a set of novel code of practices that may assist in improving the quality of research. 

\subsubsection{Quality of Quantitative Energy Modeling Studies}

A majority of the studies analyzed are quantitative energy models, thus we start our discussion by providing an overview of quality of quantitative energy models. We highlight our results per quality indicators from Table \ref{tab:QIQEM}. The following subsection highlights some of the common issues while providing a thematic overview of each of these quality indicators. We refer the reader to the supporting material\footnote{Supporting material is available at: www.github.com/ashishrsai/energy} for more detailed results. 
\begin{enumerate}
    \item \textbf{QM1:} Hardware Assumptions  \begin{enumerate}
    \item Improper hardware efficiency assumptions:
    \begin{itemize}
        \item In \cite{mora2018bitcoin}, the authors keep the power efficiency of mining hardware constant while conducting their simulation for the next 100 years. They do not provide clear reasoning behind their choice. This has been criticized by many matters arising published in Nature Climate Change (\cite{masanet2019implausible,houy2019rational,dittmar2019could}). These studies have demonstrated that this choice alone impacted \cite{mora2018bitcoin}'s results considerably, rendering their analysis provably inaccurate. 
    \end{itemize}
    \item Assuming single hardware in use:
    \begin{itemize}
        \item A more common error in many of the analyzed studies is in their decisions regarding the composition of the hardware pool, including the choice of hardware and the proportion of each machine. In their initial attempt \cite{o2014bitcoin}, the authors do not consider a mix of hardware for their analysis, they generate their estimate with the lowest (commodity) and highest (specialist) hardware alone. This likely skews their results considerably \footnote{It is worth noting that the \cite{o2014bitcoin} indicates that Bitcoin likely consumes as much electricity as Ireland without providing any calculation to support this assertion. This has already been criticized by \cite{koomey2019estimating,vranken2017sustainability}}. It is important to point out that these two estimates provide the reader with a lower and upper bound of the energy consumption of Bitcoin. Both the lower and upper bounds are likely accurate but unrealistic as not everyone on the network would operate the most or the least efficient piece of hardware. 
        \item  In another article (\cite{gallersdorfer2020energy}), we note that the estimates are based on a single hardware efficiency rather than a pool. It is not clear why the authors picked the specific hardware that they did, for instance, for Bitcoin, the authors picked Bitmain Antminer S17 Pro 53TH however they do not provide evidence or rationalization of why this is a good representative of the network. At the time of publication, more efficient hardware such as the Bitmain Antminer DR5 (34Th) was already in use. 
    \end{itemize}
    \item Filling in the missing data:
    \begin{itemize}
        \item In a more recent study by \cite{stoll2019carbon}, the authors attempt to improve the accuracy of the hardware pool by reviewing IPO filling for three major bitcoin mining hardware providers. They assume that the top three hardware providers control all of the supply despite each listed IPO acknowledging that there is about 5 to 15\% of supply beyond these three providers. The exact distribution between these three suppliers also varies considerably between the three IPOs and the article itself. We believe this limits the reliability of this model considerably. The authors additionally make many assumptions that are not based on the data provided within the IPO reports, such as the assumption in their supplementary Sheet 3.4, these assumptions lack empirical basis while being consequential to the final prediction. \\

        \textbf{Assumption made by \cite{stoll2019carbon}:} 

        \begin{enumerate}
            \item The IPO filing does not specify the sales figures per model thus the authors assume an equal distribution of sales on all available ASICs models.
            \item Post the publication of the IPO filings, the authors assume that the number of ASICs sold per month in 2018 stays constant. 
        \end{enumerate}
 
        Both of these assumptions may skew the results in favor of old hardware that might not be as efficient as newer iterations. Similar assumptions of the equal spread of device sales over a time horizon are also foundational to the work presented in \cite{de2018bitcoin}.
        
        \item  In a follow-up work, \cite{de2020bitcoin}, the author attempts to justify how old devices (specifically Antminer S9) are still the dominant mining hardware, this is supposedly backed by the evidence in Supplementary Data Sheet 1, however on inspecting the data-sheet we notice that in 2019, for H1 (first half of the year), the author has assumed (\textit{Assumption b}) that the sales of Bitmain devices are equally distributed amongst all models (including the old Antminer S9 and newer S11, etc). This assumption is not justified by any empirical evidence or rationalization. Similarly, for Q3 in 2019, the author applies \textit{Assumptions e and f}. Assumption e establishes an arbitrary distribution of sales between the three producers, this distribution is not backed by any empirical evidence. Assumption f assumes that the ratio of sales between two ASIC devices (including S9) remains the same as H1 of 2019 and is equally spread between available models.
    \end{itemize}
    \item Random Distribution of Hardware in the pool:
    \begin{itemize}
        \item Another type of error in constructing the hardware pool is a random distribution of hardware share. For instance, in \cite{mora2018bitcoin}, the authors assume a random assignment of mining hardware that leads to an equal probability that old hardware is used to mine a block as frequently as a newer, more efficient one leading to an inflated energy consumption figure.
    \end{itemize}
    
\end{enumerate}
\item \textbf{QM2:} Geographic Data
\begin{enumerate}
     \item Using mining pool location:
     \begin{itemize}
         \item Mining pools are widely used in bitcoin mining as it is often profitable to mine in a group rather than individually \cite{lewenberg2015bitcoin}. Participation in a pool can lead to an overall improvement in expected return from the mining operation. Studies such as \cite{mora2018bitcoin}, have used the IP address of a mining pool to assign a geographic location to a miner. This approach has been questioned extensively in the literature as anyone can join a mining pool irrespective of their geographic location \cite{masanet2019implausible}. This approach was also used by \cite{onat2021bitcoin}, where the study attributes the central server of a pool as the sole geographical location where all the mining takes place. This is considered provably inaccurate \cite{masanet2019implausible}.
     \end{itemize}
     \item Extrapolation of Data:
     \begin{itemize}
         \item Using Cambridge Data-set: the University of Cambridge through their annual surveys has been able to map 32\% to 37\% of Bitcoins mining power to a specific geographic location (\cite{ccaf}). This data set explicitly states that it only captures 32\% to 37\% of all the computing power in the network, despite this, in \cite{de2022revisiting}, authors have used anecdotal examples to justify their reliance on the extrapolation of the Cambridge data set to represent the whole network. The two non-academic literature examples used do not provide scientific evidence of actual distribution thus it is safe to assume that the validity of this study is considerably limited.
         \item Using public pools data: In \cite{stoll2019carbon}, the authors use the location distribution of the mining pool BTC.com and assume that it is representative of all Chinese pools, this is not backed by any reason or data. Similarly, the authors also use the location distribution of Slushpool and assume it to be representative of all European pools. They additionally treat Unknown/other pools as Chinese or European pools according to the ratio of Chinese to European pools. All of these assumptions have an impact on the accuracy of their model and their subsequent results.

     \end{itemize}
 
        \end{enumerate}
\item \textbf{QE3:} Economics
\begin{enumerate}
     \item Cost of Electricity:
     \begin{itemize}
         \item Cost assumption without empirical evidence: In their attempt, \cite{o2014bitcoin}, the authors assumed the electricity cost that may be considered high (0.10 USD per kilowatt-hour). The authors reason their choice by selecting the lowest electricity cost in the eurozone. As it is established now that a majority of mining occurs in countries with lower costs (\cite{ccaf}), this assumption may not hold true. In his work, Digiconomist assumes electricity cost to be 0.05 USD, however, their choice is not backed by any empirical data. In the University of Cambridge's Bitcoin Electricity Consumption Index, they also chose 0.05 USD as a reasonable estimate by citing ``conversations with experts" as justification however there is no evidence, such as, interview transcripts to make the claim traceable.
         
         To demonstrate the variance in the choice of electricity cost, we plot the choice from all the reviewed studies in Figure \ref{fig:COE}. It can be seen from the Figure \ref{fig:COE}, that there is a trend towards using 0.05 USD/KWh as the value since 2019. We attribute this trend to the introduction of the Cambridge index and their choice of electricity cost. 
         
         \begin{figure}
             \centering
             \includegraphics[scale=0.40]{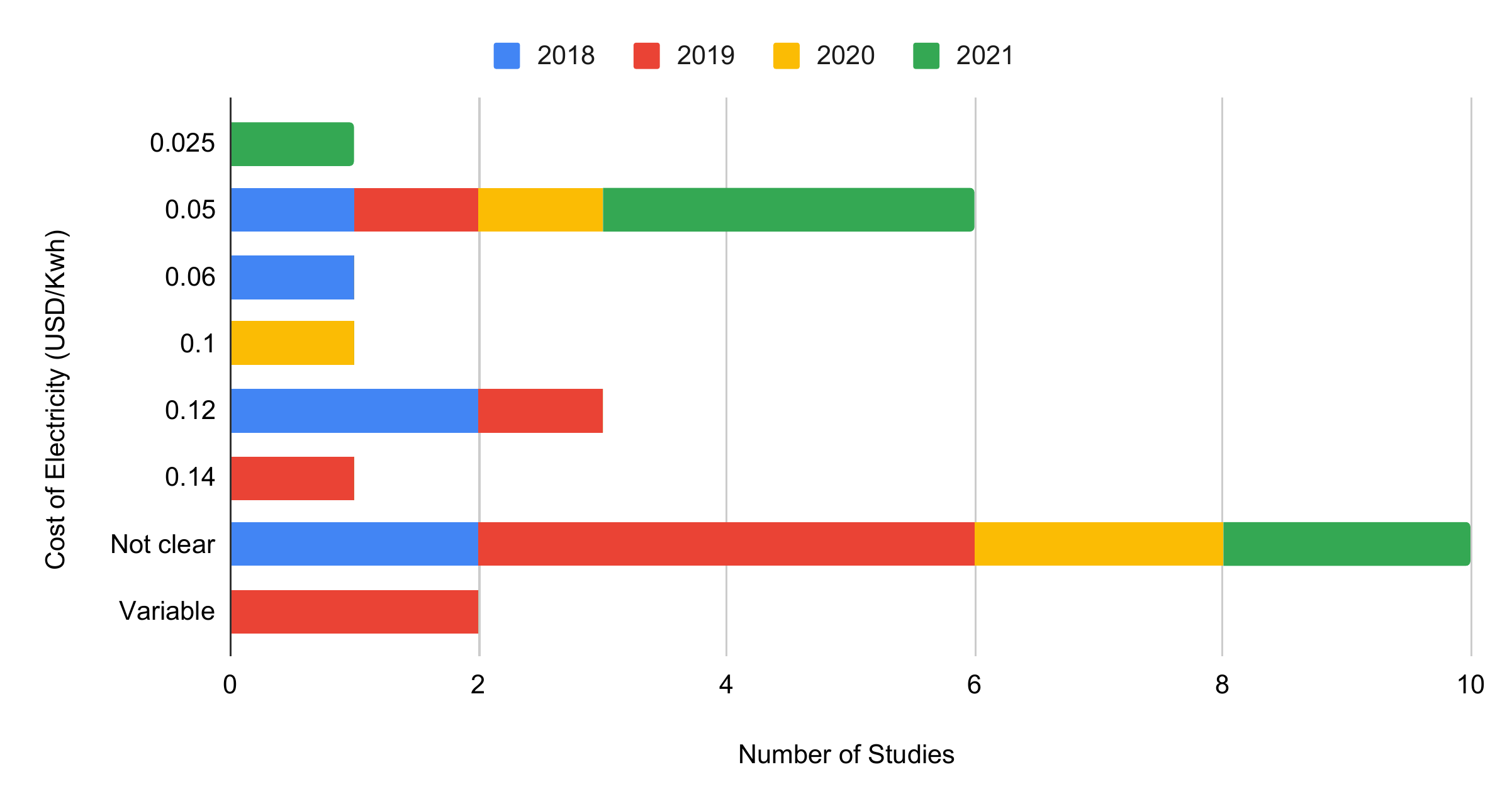}
             \caption{Choice of the cost of electricity over time}
             \label{fig:COE}
         \end{figure}
     \end{itemize}
     \item Power Usage Effectiveness (PUE):
     \begin{itemize}
         \item Not considering PUE: A number of works such as \cite{vranken2017sustainability,gallersdorfer2020energy} have excluded PUE consideration from the analysis likely limiting the validity of their results to just machines used for mining. 
         \item No justification for the selected PUE value: This is surprisingly common in studies that we examined. We believe this to be a significant limitation of the current estimates. The Cambridge Index (\cite{ccaf}) provides no traceable justification for PUE value selection. They do state that their conversations with the industry suggest a low PUE value. However, they do not provide any interview transcripts or supporting information. Similarly, \cite{de2020bitcoin} does not provide any justification for their choice of PUE. In the supporting evidence, the PUE value is based on an assumption (\textit{Assumption i}). This is also an issue with \cite{stoll2019carbon} to a lesser extent, where the authors select PUE values for small, medium, and large miners based on an interview with one miner. This we believe is very limited as their analysis demonstrates that a change of PUE from 1 to 1.3 results in about 26\% more energy consumption.
        
         \item Using hardware lifespan to estimate PUE: In \cite{de2018bitcoin}, the author states that the electricity consumption only accounts for 60\% of all revenue costs. The author attempts to justify their choice of this arbitrary number through the aforementioned assumption of the limited life cycle of 1 to 2 years. This has been widely criticised both in academic (\cite{koomey2019estimating}) and non-academic literature (\cite{bevand_2017}). 
     \end{itemize}
 \end{enumerate}
 
\item \textbf{QM4:} Social
\begin{enumerate}
  \item Hardware Lifespan:
     \begin{itemize}
         \item Non-empirical assumption: In \cite{de2018bitcoin}, the author assumes that the life cycle of Bitcoin hardware is around 1 to 2 years. The author uses Antminer S9 as an illustrative example, however, since the publication of this article in 2018, we have seen continuous use and sale of S9\footnote{This is also evident in IPO data analyzed by \cite{stoll2019carbon}.} up until the end of December 2021. This is a lifespan closer to 4 to 5 years\footnote{The IPO filling and other sources online tend not to differentiate between the different iterations of the same hardware. For instance, the Antminer S9 sold in 2019 might have a different performance and energy profile when compared to Antminer S9 sold in 2017. This lack of clarity might undermine the suggested 4 to 5 years of lifespan.} as opposed to the 1 to the 2-year estimate given by the author. This work served as the foundation of follow-up work on carbon emission of Bitcoin \cite{k2019}, we believe that the reliance on \cite{de2018bitcoin}'s estimate is questionable.
     \end{itemize}
\end{enumerate}
\item \textbf{QM5:} Carbon Emission Data

\begin{enumerate}
    
     \item Applying old energy mix and carbon intensity data to current or future predictions:
     \begin{itemize}
         \item In \cite{mora2018bitcoin}, the authors used 2014 carbon intensity data on 2017 analysis, this led to inaccurate estimates as pointed out by \cite{masanet2019implausible}. Similarly, in \cite{de2022revisiting}, the authors rely on the data for energy mix from 2019 and apply it to 2021. They do acknowledge that they are using old data due to the lack of availability of new data-set however they do not explain how it impacts the reliability of their results, making it difficult for the reader to assess the reliability of their study. We also note that in \cite{onat2021bitcoin}, the authors used data from November 2018, and applied it to all estimates in 2015 up to 2020. 
     \end{itemize}
  
\end{enumerate}
\item \textbf{QM6:} Time Resolution
\begin{enumerate}
    \item Not reporting time of measurement/prediction/analysis
    \begin{itemize}
        \item A substantial chunk of the analyzed studies (32\%) does not document the time of their analysis or measurement. The lack of a time resolution makes it difficult to understand the reliability of the model. For instance, without explicitly mentioning the time of data collection for Bitcoin mining hardware, it is difficult to understand if the model captures appropriate hardware at the time. This is also true for the case of carbon emission data as illustrated above.  
    \end{itemize}
    \item Not documenting the evolution of model parameters
\begin{itemize}
    \item Real-time indexes such as CCAF and Digiconomist evolve over time due to the changes in their underlying data or changes in assumptions. Documenting these changes can assist in understanding the evolution of these models, for instance, CCAF provides a change log. However, the Digiconomist index does not seem to have a change log documenting the changes over time.
\end{itemize}
\end{enumerate}
\end{enumerate}

\subsubsection{Quality of other Research Methods}
As is evident from our analysis presented above due to the infant nature of this research domain, there is a considerable number of assumptions each of which may impact the accuracy of the results. Unlike the quantitative models that we have discussed so far, it is relatively easy to assess and improve the rigor of other methodologies due to the presence of vast literature on best practices associated with each of these methodologies. In this subsection, we provide an overview of our analysis. As before, we advise the reader to refer to the supporting evidence for an in-depth analysis of the results. 

\begin{figure}
    \centering
    \includegraphics[scale=0.4]{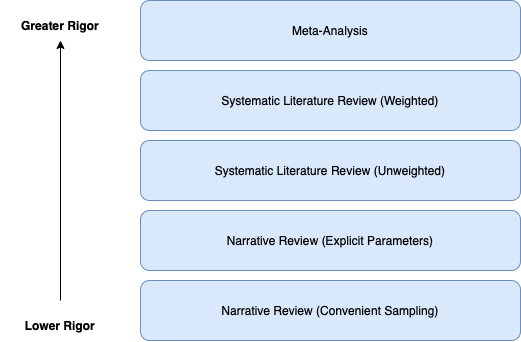}
    \caption{LR1: Type of review method and associated rigor (Figure adopted from \cite{sovacool2018promoting,khan2003five,huebner2017we})}
    \label{fig:litReview}
\end{figure}
\begin{enumerate}
    \item Literature Review: A majority of the reviews (13 out of 16) were narrative reviews that according to \cite{sovacool2018promoting} are not as rigorous as other forms of reviews\footnote{See also Figure \ref{fig:litReview}} (LR1). Both systematic literature reviews and meta-reviews were not prominent. Only three of the analyzed studies had documented the search and inclusion criteria making it possible to replicate the studies and also independently estimate the coverage of the review (LR2). We report a similar trend in documenting the search databases and sampling process as only 3 studies report these (LR3 \& LR4). This is a major limitation in literature reviews as it is difficult to assess the coverage and quality of the analysis without the ability to independently replicate it.
    \begin{figure}[]
    \centering
    \includegraphics[scale=0.6]{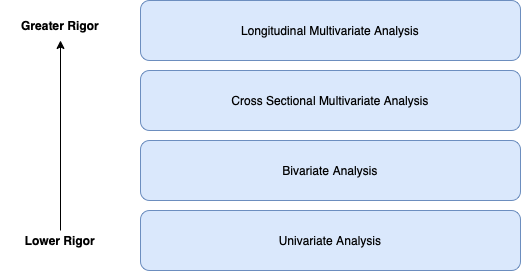}
    \caption{DA1: Data analysis and statistics methods (Figure adopted from \cite{tabachnick2007using,sovacool2018promoting,denis2015applied})}
    \label{fig:litReview}
\end{figure}
    \item Data Analysis and Statistics: Unlike Literature reviews which reported that a majority of the studies employed less rigorous research methods, in data analysis and statistics, a majority of the studies (7 out of 13) employed high rigor multivariate longitudinal analysis. However, only 4 studies documented the hypothesis clearly making it difficult to assess the results of the analysis. Despite the use of more rigorous research methods, these studies suffer a big limitation in acknowledging the practical significance of their results. For instance, \cite{jana2022taming} attempts to use a deep neural network to predict electronic waste generation from Bitcoin however it is not clear if their results are a good representation of the actual Bitcoin ecosystem or how the predictions could be used in a practical context. 
  \item Case Studies: Only a small number of analyzed studies (6) used a case study approach. These studies employed a diverse range of selection mechanisms, illustrative being the most popular. Our analysis suggests that it is difficult to assess the appropriateness of a selected case due to the complicated nature of cryptocurrencies. For instance, in \cite{liu2021bitcoin}, the authors select an influential case of mining in Sichuan and Xinjiang in China to understand the incentives associated with mining there however it could be argued that the case selection should have accounted for the seasonality of Bitcoin mining\footnote{It is also worth noting that since the imposition of the Chinese ban on mining activities in China in 2021, it has become increasingly difficult to perform mining in the Sichuan and Xinjiang.}. A bigger issue with these studies is the lack of clear boundaries, for example, in \cite{gundaboina2021energy}, the authors examine the energy and resource consumption of several hashing algorithms in Dogecoin by using a selective piece of hardware without explicitly describing the bounds of the analysis. In a majority of these analyses (1 out of 6) it is unclear what the dependent and independent variables are. 
  \item Experiments: Experiments are increasingly popular in Proof-of-Stake cryptocurrencies as it is not difficult to generalize the results from these experiments to the whole network as they mostly rely on off-the-shelf hardware \cite{platt2021energy}. However, it is still important that the selected hardware/s for experiments are representative of the network's hardware composition. Out of 4 analyzed experiments, only one had used a large set of hardware devices to conduct the experiments. For instance, \cite{roma2020energy} performs an analysis of the Ripple network however they only use a single piece of hardware with a specific CPU without discussing how representative the hardware is of the whole network. All of these studies do not account for the settings of the experiment such as geographic variation. This may play a significant role when calculating the real-world performance of these devices. For instance, in \cite{li2019energy} the authors conduct an experiment in an isolated room where they remove operational externalities such as heat, this may not be a good representation of the operational state of mining hardware. 
\end{enumerate}

  In the previous subsection, we provided a  breakdown of common issues in energy consumption studies, highlighting specific instances from the shortlisted studies. In this subsection, we provide a summary of issues for two of the most popular models to better contextualize our findings.

\subsection{Limitation of popular approaches}

\subsubsection*{Cambridge Bitcoin Electricity Consumption Index (CBECI)}
  Based on our analysis, we consider CBECI to be one of the more carefully done analyses of Bitcoin's electricity consumption, however, we would still caution policymakers and researchers from basing their decisions solely on this metric as CBECI also has known limitations and issues. 
  
  The most prominent of these issues is the choice to use nonce analysis methodology by Coin Metrics (\cite{coinmetrics_2020}) to improve their hardware pool distribution\footnote{CBECI in their recent update removed the CoinMetrics API disucssed here, however this discussion still highlights that CBECI can also suffer from seemingly trivial issues such as reliance on an unvalidated metholdogy.}. \cite{ccaf} utilizes a basket of hardware devices including two Antminer devices S7 and S9. To estimate the share of hashing power originating from Antminer S7 and S9 line devices, \cite{ccaf} relies on the data provided by \cite{coinmetrics_2020}. On inspection of the nonce analysis methodology, we note that this methodology overestimates the share of Antminer S7 considerably\footnote{This flaw in the methodology was also flagged by Digiconomist in his blog}. Their analysis suggested up to 4 million active Antminer S7s which is significantly higher than the reported values from the IPO of Antminer. We also note that the hardware pool used by \cite{ccaf} does not contain an up-to-date list of all ASIC miners. 
  
  Another critique of the CBECI is regarding their choice of PUE value, which is significantly better than Google Data Centers\footnote{See https://www.google.com/about/datacenters/efficiency/.}. CBECI backs their choice by claiming that, based on interviews and discussions with experts, this is realistic. However, due to the lack of supporting evidence such as interview transcripts, we are unable to independently verify these claims.
  
  As indicated in QM 6, CBECI does provide a log of changes made to their model however they do not document the exact parameters and their specific values used in each iteration of the model. This makes it difficult to clearly understand the evolution of the index over time as the old predictions made by CBECI may also change retrospectively with newer model changes. 
  
  \subsubsection*{Digiconomist's Bitcoin Energy Consumption Index}
  
  The approach used by the Bitcoin Energy Consumption Index has been widely debated both in academia (\cite{koomey2019estimating}) and non-academic (\cite{bevand_2017}) literature. Our review suggests that the foundational approach used by 
  Digiconomist as outlined in \cite{de2018bitcoin} is of questionable scientific rigor. We specifically question the choices made regarding the life span of the hardware and the subsequent calculation of the 60\% ratio of electricity cost in miners' revenue. Unlike the CBECI model, the Digiconomist model does not provide a clear change log of the model evaluation.
  
  We suggest that this work should be considered with caution. As indicated in the earlier section, the follow-up work \cite{de2022revisiting} of Digiconomist suffers from some issues as well, most noticeably the assumptions made by the author regarding the equal distribution of hardware sales.

\section{Code of Practices \label{COP}}

As alluded to in the introduction section, one of the main contributions of this review is the preparation of novel research guidelines that could assist in improving the rigor of research in blockchain energy sciences. In the previous section, we highlighted some of the common issues in terms of the scientific quality of the reviewed studies. 

The primary recommendation is to avoid those pitfalls however we acknowledge that this might be difficult to accomplish due to the lack of reliable datasets and difficulty in acquisition of newer data. In case it is difficult to avoid any of the above-mentioned pitfalls, we give some specific recommendations on how the researcher could present their research to avoid giving inaccurate predictions, measurements, or estimates.

In this section, we propose novel codes of practices segmented into three broad categories: basic research design, quantitative energy modeling, and other research recommendations. 
\subsection{Basic Research Design}
We have three broad recommendations when designing and executing a blockchain energy study, these guidelines are intended for both academic and non-academic studies. 

\subsubsection{Explicit Research Methodology}
\begin{figure}
    \centering
    \includegraphics[scale=0.29]{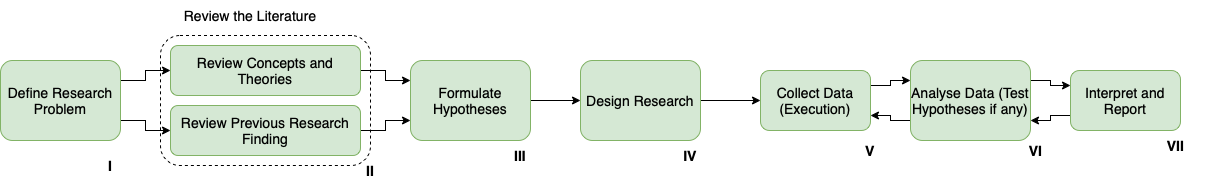}
    \caption{Research Design as proposed by \cite{kothari2004research}}
    \label{fig:RM}
\end{figure}

Any study measuring, estimating, or predicting the energy or environmental footprint of a crypto-asset should be designed to be easily reproduced. To this end, our first recommendation is to explicitly document the methodology employed in the study. We recommend that the research be documented using the 7-step model proposed by \cite{kothari2004research} as illustrated in Figure \ref{fig:RM}. We provide a brief overview of these 7 steps in the following text however it is recommended that the reader refers to the detailed description as provided by \cite{kothari2004research}.

\begin{enumerate}
    \item Clear research problem: The research problem addressed by the project must be sufficiently documented in a clear manner. This assists the reader understand the project's intended goal without any ambiguity.
    \item Review the literature: As discussed in the results section, due to the lack of coherence in this newly developing space, a lot of research development is done in small silos resulting in likely slow progress. Thus we recommend that newer models review existing literature thoroughly. To that end, our survey provides the reader with an extensive list of relevant academic and non-academic literature. We also document progress made by non-academic source\footnote{This document is public, we hope researchers and practitioners would refine and update it as needed, it can be accessed at: https://bit.ly/3csZLg9}. 
    \item Development of a hypothesis: Ideally after doing an extensive literature review, the researcher should have a starting point within the literature, this could be an existing model that the researcher intends to build upon or a research methodology that the researcher intends to adopt. Before the execution of their research methodology, it is important to clearly state the working hypothesis or hypotheses. The hypothesis should be as concise as possible to allow testing through the research at the end. 
    \item Research Method: The researcher is expected to succinctly document the steps taken by them to test a hypothesis to address the research problem. This should be documented in form of a conceptual structure of the research conducted. In the preceding section, we have provided an overview of different research methodologies adopted in order to answer a specific type of research question. For instance, if the researcher wants to get an overview of the carbon emission of a popular cryptocurrency, they might benefit from first doing a literature review, ideally by following a more rigorous method within literature reviews. If the research question at the end requires data collection\footnote{Based on our experience, this is required when conducting a quantitative energy study.}, it is important to do the sampling of the data appropriately. For instance, if it is not feasible to collect data on all mining hardware employed in a specific PoW network, an appropriate selection mechanism should be employed rather than randomly sampling a selective set. 
    \item Data Collection: There are many ways of collecting data for energy modeling, documenting each of these methods is beyond the scope of this review. We refer the reader to \cite{kothari2004research} for more instruction. However broadly speaking it is important to test the reliability of the data collected either through a primary source or a secondary source. We provide more specific instruction on this for quantitative energy modeling in a subsequent section. 
    \item Analysis of the data: Analyzing the data is crucial for reporting at this stage as the analysis can skew the interpretation considerably. For instance, an analysis conducted only in summer might overestimate the overall energy consumption of mining in China as during rainy reasoning the mining moves to hydroelectric power\footnote{It is worth noting that this phenomenon likely does not occur anymore due to the ban of mining in China however similar patterns might occur in other geographical areas.}. Similarly, solar powered plants do not generate electricity during night, wind mills only generate power if there is wind. The analysis should account for extremities as well as known limitations as outlined in our results section. At this stage, it is also crucial to test any hypothesis proposed in step 3. We provide more specific instructions on improving the analysis and highlighting the limitations based on research methods in the following subsection 5.3 and 5.4. 
    \item Interpret and report: It is important to contextualize the data generated through the analysis. For instance, in \cite{de2022revisiting}'s work, the author uses a small (34-40\%) geographic dataset from Cambridge and presents the results for the whole of the network justifying it through a selective set of qualitative non-academic literature sources. Our recommendation here is that the analysis of the data generated through these models or accumulated through a review should always be compounded with confidence intervals or sensitivity analysis to make clear the impact of assumptions or lack of data on the models' performance.
\end{enumerate}
\subsection{Sharing the Data and Source Code}
We identified accessing the data and source code as one of the main limitations to independently reproduce the results from a large set of studies we analyzed. To account for this limitation, we present two sets of recommendations in line with \cite{stodden2010reproducible}:
\begin{enumerate}
    \item Adhere to the following 3 long-term goals proposed by \cite{stodden2010reproducible}:
    \begin{enumerate}
        \item \textit{Use version-control system:} The dataset should be version-controlled, this is specifically important for live indexes such as Digiconomist and Cambridge. 
        \item \textit{Provide standardized citations for data:} in order to build upon existing datasets, the authors should provide standarised citations for the datasets. For instance, the authors in \cite{zade2019bitcoin} provide a citable source of the dataset for mining hardware allowing others to build on top of it. 
        \item \textit{Describe data using standardized terminology and ontologies:} One of the biggest hurdles in building upon an existing dataset is the way data is reported by different studies. We argue that there is a need for standardization in order to incentivize improving existing datasets rather than replicating the work again. We suggest adhering to the template used by Cambridge for documenting hardware used for Bitcoin \footnote{See http://sha256.cbeci.org}. 
    \end{enumerate}
    \item We strongly recommend that if a study is using data from an existing work, the authors supply data or a link to the data to promote transparency and verifiability of their analysis. Ideally, this should be done by using the above-listed guidelines.
    \item Hardware distribution and location data should contain collection and validation steps.
    \item If location data is used, we recommend that it be assessed for any seasonality patterns that might exist. 
\end{enumerate}

Another big limitation as outlined in the results section was the lack of source code. Source code for energy analysis often takes two forms: mathematical models implemented using a programming language or excel sheets. We provide two guidelines for sharing source code to account for both methods: 
\begin{enumerate}
    \item Excel Sheets: Authors should provide details on Information Quality (IQ) and Data Quality (DQ) proposed by European Spreadsheet Risks Interest Group \cite{o2008information}.
    \item Source Code: Similar to the long-term recommendations for sharing the data, \cite{stodden2010reproducible} suggests that the source code should be version controlled and should contain test routines that allow independent testing of the source code.
\end{enumerate}

\subsection{Quantitative Energy Modeling}
For conducting quantitative energy modeling in a blockchain context, we provide 6 guidelines in addition to the basic research guidelines proposed above. It is also recommended that the common issues outlined in the last section are also avoided. 

\begin{enumerate}
    \item Traceable \& verifiable justification for hardware assumptions: One of the biggest issues with quantitative energy modeling studies is the lack of evidence for assumptions made by studies on the hardware. We specifically suggest that the authors:
    \begin{enumerate}
        \item  State the assumptions within the text and not in supporting material
	\item Add sensitivity analysis or confidence intervals when filling in missing data. 
    \end{enumerate}
    \item Traceable \& verifiable justification for economic assumption: Similar to assumptions about the hardware in use, in economic models, it is vital that the authors back their assumptions with evidence. We specifically suggest that the authors: \begin{enumerate}
        \item  Should include both capital and operational costs for different agent types (small, medium, and large)
        \item Cost of electricity should be as granular as possible, if data is missing, use location-based metric.
        \item Hardware lifespan assumption should be validated using real-world data
    \end{enumerate}
	\item Using or collecting geographic data: The geographic data plays a crucial role in modeling the environmental footprint of these crypto-assets. We recommend that the authors: 
	\begin{enumerate}
	    \item Should avoid using mining pool IP address, if used, it should be accompanied by sensitivity analysis or appropriate confidence intervals.
	   \item Should avoid using non-academic literature, unvalidated sources for location.
	   \item Should include the date of data collection.
	   \item Should not extrapolate location data, if done, should be accompanied by sensitivity analysis or confidence intervals
	\end{enumerate}
	\item PUE value should be based on empirical evidence. The modeling should also include different types of agents (small, medium, and large) if possible. 
	\item Avoid unreliable sources of data such as proven faulty studies  \cite{mora2018bitcoin})
	\item The authors should also avoid using improper units of comparison. For instance, Bitcoin and Ethereum do not consume electricity per unit of transaction but per block \cite{sedlmeir2020energy}. Comparing per transaction electricity consumption of Bitcoin or Ethereum may be inaccurate or misleading according to \cite{sedlmeir2020energy}.
\item The authors should account for the time resolution of the data and model parameters. We suggest using time stamps for all the data used in the model and the model itself if it evolves over time.
\end{enumerate}

\subsection{Other Research Methodologies}
Similar to quantitative energy modeling, the prime recommendation here is to avoid the common pitfalls outlined in the result section and follow the basic research design guidelines proposed above. We have documented our guidelines in more detail on the research repository located at https://cryptocop.vercel.app. We strongly recommend that the reader refers to the repository to see the in-depth guidelines for these methodologies. Here we provide an overview of the specific suggestions for other research methodologies: 
\begin{enumerate}
    \item Literature Review: We strongly recommend that the authors adopt more rigorous forms of literature reviews such as meta-review or systematic literature reviews. It is advised that the authors follow the guidelines proposed by \cite{kitchenham2007guidelines} to improve the coverage and reliability of the review. 
    \item Data Analysis and Statistics: For data analysis and statistics our prime recommendation is that the authors should avoid oversimplified analysis and account for the practical significance of their results. 
    \item Case Studies: As indicated in the results section, the selection of the case study should be appropriate. We recommend that the authors follow the guidelines put forth by \cite{sovacool2018promoting} for conducting a case study analysis. 
    \item Experiments: Our main recommendation for conducting experiments is to choose an appropriately large sample size. 
    
\end{enumerate}

\section{Discussion}\label{discussion}
There is a clear trend that both academic and non-academic researchers have paid increasing attention to the energy and environmental footprint of blockchain technologies in the recent years. In academia, this trend is predominately focused on accurately understanding and predicting the electricity consumption of popular cryptocurrencies such as Bitcoin and Ethereum. There is also a smaller fraction of academic literature that focuses on other smaller cryptocurrencies such as IOTA and Ripple. 

A majority of the academic research in this domain is either purely applied or uses inspired basic research with little to no focus on pure theory development\footnote{We refer the reader to \cite{sovacool2018promoting} for an in-depth disucssion of different types of research in social energy sciences.} or refinement. This was also reflected through our quality indicator BR2, where we reported a large chunk (74\%) of academic literature does not build upon existing knowledge. 

Unlike the academic literature, non-academic sources had a more broad focus on different types of cryptocurrencies. However, most of the research in the non-academic sphere is sponsored as we could anticipate. In terms of research design, we notice a trend of more applied research with little to no focus on theory building.

\subsection{Current Problems and Potential for Improvement}

We do support the efforts to model, estimate, and predict the energy and environmental footprint of these crypto-assets. Our research in this article clearly documents the evolution of the field. We believe this evolution will likely keep spurring more discussion and public debate on this worthwhile topic. However, we also note that due to the focus on unreliable results may blur or even misguide this discussion. 

Our review and analysis make clear that there are two broad problems:

\begin{enumerate}
    \item The most prominent one is the lack of rigor in many published studies. We have highlighted some specific instances of this throughout our results section while providing an abstract overview of the whole field where appropriate. This lack of rigor might suggest that the estimates and predictions on energy consumption and environmental footprint are likely questionable.
    \item Our analysis also shows that the current system for publishing scientific/academic literature is lacking in appropriate scrutiny. The publishers, journal editors, program committees of conferences, and reviewers are responsible for ensuring the quality of publications.  We show that quality is below what one may expect. However, we also acknowledge that it is always easy to comment when looking back. As a field evolves, prior errors become evidently visible.
\end{enumerate}

Overall, based on our analysis using the quality indicators proposed in Section \ref{3methodology}, we report that most of the models used for energy and environmental footprint estimation suffer many known flaws that limit their reliability. We argue that more empirical data has to be collected before these estimates can be considered accurate and scientifically rigorous for policy decisions. 
  
 Many of the reviewed studies suffer from trivial issues such as unsubstantiated claims or assumptions. This is uncharacteristic of a mature scientific domain, however as this field is in the early stage, these measures are likely to see further refinement both in terms of their reliability and accuracy. We identify three potential avenues for improvement in the state of this field: standardization, quality assessment, and provision for more data collection. 
 
 \begin{enumerate}
     \item Standardisation: Before we can start meaningfully discussing how to improve the quality of these models, the field must agree on the scope and definitions of these models. For instance, a model designed to estimate the electricity consumption of a PoW-based cryptocurrency should describe all the fundamental building blocks such as hardware efficiency, PUE value, and cost of electricity in a transparent manner. The measurements generated by these models should also include a clear timestamp. The data used by these models should also adhere to the nomenclature\footnote{We suggest that the field begins with adopting the nomenclature used by Cambridge dataset at http://sha256.cbeci.org/}. This standardization not only helps in comparing the results of different models but also transparently understanding the assumptions made by each of these models. It may also promote more collaboration by building upon existing open datasets rather than re-building the same datasets again. 
     \item Quality Assessment: This study proposes an initial quality assessment framework in the form of numerous quality indicators. However, we believe this to be only a starting iteration of the framework. We have attempted to logically split the framework based on different research methodologies. This framework can easily be extended to account for different consensus mechanisms. We must be able to discuss the robustness of a given model transparently in order to make policy choices based on the outcome. 
     
     One of the main recommendations of our model is the use of sensitivity analysis and confidence intervals in order to communicate results. This assists in ensuring that the results are better contextualized. Another important outcome of our survey is the development of novel code of practices. We believe that a document structure that adheres to the guidelines proposed in subsection 5.1 would allow for an easy evaluation of the scientific quality of the report. 
     \item Provision for more data collection: One of the prime issues associated with models for electricity and environmental footprint in this domain is the lack of real-world data regarding the machines in use or their geographic location. We believe that this problem needs to be addressed by both private and public stakeholders in this ecosystem. Private mining pool operators should attempt to document and validate their energy consumption and use of renewable claims. Whereas public bodies should attempt to design regulatory frameworks to either incentivize or impose reporting of energy use. This can assist in validating the models that we already have in place while also promoting more rigorous reporting. 
    
 \end{enumerate}


\section{Conclusion\label{conclusion}}
In this paper, we conduct a systematic literature review to provide a summary of research done on the energy and environmental footprint of blockchain-based technologies by both academic and non-academic sources. We contextualize our findings by analyzing the robustness of these studies pointing out common issues while also highlighting potential avenues of either fixing the issues or presenting the results better to account for the limitations. 

Given the significant growth of blockchain technologies in recent years and their potential impact on the environment, we believe this study to be pivotal in encouraging a constructive discussion on the reliability of the models used to measure and in some cases offset the carbon emissions generated through the use of blockchain. This refined understanding of the common issues faced by studies focusing on energy and environmental footprint allows us to generate a set of recommendations in the form of code of practices that may improve the overall quality of these models. 

\subsection{Contribution}

We systematically review the literature from both academic and non-academic sources. This allows us to build a large corpus of models used to measure, estimate or predict the energy or environmental impact of blockchain technologies. We not only documented these models but also attempted to traceably discuss the potential limitations of these models. We do this by using the quality indicators from cognizant fields of social energy sciences, information systems, and computer science. 

Our analysis suggests that a majority of these studies lack the scientific rigor expected from a mature scientific field. We make specific suggestions regarding the reuse of existing theory and datasets to promote more cohesive research and hopefully iteratively improve these models. In order to measure or contextualize the quality of the research, we adopt and refine the codes of practices from \cite{sovacool2018promoting}.

Through our review, we have documented the substantial progress made by blockchain energy researchers in obtaining novel datasets and constructing useful models under the constraints of a decentralized system. We strongly support and encourage the development of these models. We believe this article provide a an enumerated list of common flaws to avoid while working on blockchain energy model.  

For quantitative energy modeling, we provide a list of common issues and identify instances from popular academic and non-academic sources that suffer these limitations. Our intention here is to highlight that even the most popular studies might be limited in terms of their validity. To this end, we present an overview of some of the issues present in both the Cambridge and Digiconimist models. In addition to the issues of quantitative energy modeling, we also highlight the common problems with other research methodologies and how they can be avoided through the adherence to best practices from cognizant disciplines. 

The main intention of our work is to promote rigor in blockchain energy sciences. To this end, we develop a set of guidelines in the form of code of practices that can be used by both academic and non-academic researchers. We believe adherence to these code of practices will not only ensure that common issues and pitfalls are avoided but also help improve the quality of the model and the accompanying report. Our code of practices are also grounded in the needs of the field specifically in terms of building on existing knowledge and also better contextualizing results for policy decisions. 

To conclude we strongly believe that this work paves the way for a constructing discussion on the topic of energy and environmental footprint by advocating for a common vocabulary while providing tools to compare different models and understand their reliability. We also believe that the code of practice will promote traceability, building upon existing work and better contextualizing the results. We  note the quality framework developed in this work is an initial framework that is intended to be built upon and iteratively refined as the knowledge around this topic improves.

\subsection{Threats to Validity}
Our analysis despite adherence to the best practices from the literature might suffer from limitations. The primary limitation is the nature of the search for both non-academic and academic literature. The terms used for search are intentionally kept vague to ensure high coverage however they return a large number of both relevant and irrelevant articles that we then filter through the title and abstract filtration process. This process might have omitted some relevant articles as ensuring high reliability of filtration is difficult in a large dataset. To limit this potential issue, we perform cross-validation and obtain a reliable result suggesting that the filtration process is reliable and may be replicated independently. 

The selection of non-academic literature was particularly challenging due to the vastness of the cryptocurrency domain. We restrict our focus to a set of top 200 cryptocurrencies and use similar keywords as the academic literature. However, this focus on a subset of cryptocurrencies might have led to the omission of some other methodologies. To account for this limitation, we also include all the signatories and supports of the crypto climate accord. 

Another potential source of limitation to our work is the selection of quality indicators from the work of \cite{sovacool2018promoting}. Due to the fundamentally different nature of crypto climate research, not all the recommendations from \cite{sovacool2018promoting} apply to all the models we reviewed. To account for this, we combine these guidelines with the work of \cite{lei2021best}. We additionally also proposed our quality assessment framework as an initial step in the development of a more robust quality assessment framework. We leave it up to future work to iteratively refine and improve this model. 

\subsection{Future Work}
The work presented in this article provides an initial set of quality indicators that could benefit from refinement specifically for non-PoW-type cryptocurrencies. As a next step, we wish to develop a number of flavors of these quality indicators for Proof-of-Stake and other popular consensus mechanisms. 

We also intend to make our code of practices more accessible to researchers and practitioners by developing a web application that provides a checklist and points out common flaws and potential avenues for fixing them. To this end, we have already designed a primitive version of these guidelines\footnote{These guidelines can be viewed at https://cryptocop.vercel.app}, however, we intend to refine it further to improve user experience. 

One of the intentions of this work is to allow researchers to revisit their existing models and refine them to avoid any of the issues pointed out in this work. To this end, we wish to refine the model developed and used by \cite{vranken2017sustainability} while adhering to our codes of practices.


\section*{Acknowledgments}

\
\bibliographystyle{unsrt}  
\bibliography{references}  

\begin{thebibliography}{10}

\bibitem{orgerie2014survey}
Anne-Cecile Orgerie, Marcos Dias~de Assuncao, and Laurent Lefevre.
\newblock A survey on techniques for improving the energy efficiency of
  large-scale distributed systems.
\newblock {\em ACM Computing Surveys (CSUR)}, 46(4):1--31, 2014.

\bibitem{stephens}
Stephens, Aug 2000.

\bibitem{koomey2008turning}
Jon Koomey.
\newblock {\em Turning numbers into knowledge: Mastering the art of problem
  solving}.
\newblock Analytics Press, 2008.

\bibitem{koomey2002sorry}
Jonathan~G Koomey, Chris Calwell, Skip Laitner, Jane Thornton, Richard~E Brown,
  Joseph~H Eto, Carrie Webber, and Cathy Cullicott.
\newblock Sorry, wrong number: The use and misuse of numerical facts in
  analysis and media reporting of energy issues.
\newblock {\em Annual review of energy and the environment}, 27(1):119--158,
  2002.

\bibitem{sedlmeir2020energy}
Johannes Sedlmeir, Hans~Ulrich Buhl, Gilbert Fridgen, and Robert Keller.
\newblock The energy consumption of blockchain technology: beyond myth.
\newblock {\em Business \& Information Systems Engineering}, 62(6):599--608,
  2020.

\bibitem{zheng2018blockchain}
Zibin Zheng, Shaoan Xie, Hong-Ning Dai, Xiangping Chen, and Huaimin Wang.
\newblock Blockchain challenges and opportunities: A survey.
\newblock {\em International Journal of Web and Grid Services}, 14(4):352--375,
  2018.

\bibitem{bitcoinvisuals}
bitcoinvisuals.com.
\newblock Block reward per block chart, 2022.

\bibitem{lei2021best}
Nuoa Lei, Eric Masanet, and Jonathan Koomey.
\newblock Best practices for analyzing the direct energy use of blockchain
  technology systems: Review and policy recommendations.
\newblock {\em Energy Policy}, 156:112422, 2021.

\bibitem{bevand_2017}
Marc Bevand.
\newblock Serious faults in digiconomist's bitcoin energy consumption index,
  2017.

\bibitem{masanet2019implausible}
Eric Masanet, Arman Shehabi, Nuoa Lei, Harald Vranken, Jonathan Koomey, and
  Jens Malmodin.
\newblock Implausible projections overestimate near-term bitcoin co2 emissions.
\newblock {\em Nature Climate Change}, 9(9):653--654, 2019.

\bibitem{koomey2019estimating}
Jonathan Koomey.
\newblock Estimating bitcoin electricity use: A beginner’s guide.
\newblock {\em May, Coin Center Report, https://www. coincenter.
  org/app/uploads/2020/05/estimating-bitcoinelectricity-use. pdf}, 2019.

\bibitem{mora2018bitcoin}
Camilo Mora, Randi~L Rollins, Katie Taladay, Michael~B Kantar, Mason~K Chock,
  Mio Shimada, and Erik~C Franklin.
\newblock Bitcoin emissions alone could push global warming above 2 c.
\newblock {\em Nature Climate Change}, 8(11):931--933, 2018.

\bibitem{houy2019rational}
Nicolas Houy.
\newblock Rational mining limits bitcoin emissions.
\newblock {\em Nature Climate Change}, 9(9):655--655, 2019.

\bibitem{dittmar2019could}
Lars Dittmar and Aaron Praktiknjo.
\newblock Could bitcoin emissions push global warming above 2 c?
\newblock {\em Nature Climate Change}, 9(9):656--657, 2019.

\bibitem{sovacool2018promoting}
Benjamin~K Sovacool, Jonn Axsen, and Steve Sorrell.
\newblock Promoting novelty, rigor, and style in energy social science: Towards
  codes of practice for appropriate methods and research design.
\newblock {\em Energy Research \& Social Science}, 45:12--42, 2018.

\bibitem{nakamoto2008bitcoin}
Satoshi Nakamoto.
\newblock Bitcoin: A peer-to-peer electronic cash system.
\newblock {\em Decentralized Business Review}, page 21260, 2008.

\bibitem{kitchenham2007guidelines}
Barbara Kitchenham and Stuart Charters.
\newblock Guidelines for performing systematic literature reviews in software
  engineering.
\newblock 2007.

\bibitem{de2021bitcoin}
Alex De~Vries and Christian Stoll.
\newblock Bitcoin's growing e-waste problem.
\newblock {\em Resources, Conservation and Recycling}, 175:105901, 2021.

\bibitem{sai2021taxonomy}
Ashish~Rajendra Sai, Jim Buckley, Brian Fitzgerald, and Andrew Le~Gear.
\newblock Taxonomy of centralization in public blockchain systems: A systematic
  literature review.
\newblock {\em Information Processing \& Management}, 58(4):102584, 2021.

\bibitem{bevand}
Marc Bevand.

\bibitem{brady2013case}
Gemma~A Brady, Nikil Kapur, Jonathan~L Summers, and Harvey~M Thompson.
\newblock A case study and critical assessment in calculating power usage
  effectiveness for a data centre.
\newblock {\em Energy Conversion and Management}, 76:155--161, 2013.

\bibitem{ccaf}
CCAF.
\newblock Cambridge bitcoin electricity consumption index (cbeci), 2022.

\bibitem{vranken2017sustainability}
Harald Vranken.
\newblock Sustainability of bitcoin and blockchains.
\newblock {\em Current opinion in environmental sustainability}, 28:1--9, 2017.

\bibitem{de2018bitcoin}
Alex De~Vries.
\newblock Bitcoin's growing energy problem.
\newblock {\em Joule}, 2(5):801--805, 2018.

\bibitem{o2014bitcoin}
Karl~J O'Dwyer and David Malone.
\newblock Bitcoin mining and its energy footprint.
\newblock 2014.

\bibitem{k2019}
Susanne Kohler and Massimo Pizzol.
\newblock Life cycle assessment of bitcoin mining.
\newblock {\em Environmental science \& technology}, 53(23), 2019.

\bibitem{johnson_pingali}
Marc Johnson and Sahithi Pingali.
\newblock Guidance for accounting and reporting electricity use and carbon
  emission from cryptocurrencies.

\bibitem{huckle2016socialism}
Steve Huckle and Martin White.
\newblock Socialism and the blockchain.
\newblock {\em Future Internet}, 8(4):49, 2016.

\bibitem{fleiss1973equivalence}
Joseph~L Fleiss and Jacob Cohen.
\newblock The equivalence of weighted kappa and the intraclass correlation
  coefficient as measures of reliability.
\newblock {\em Educational and psychological measurement}, 33(3):613--619,
  1973.

\bibitem{sim2005kappa}
Julius Sim and Chris~C Wright.
\newblock The kappa statistic in reliability studies: use, interpretation, and
  sample size requirements.
\newblock {\em Physical therapy}, 85(3):257--268, 2005.

\bibitem{kothari2004research}
Chakravanti~Rajagopalachari Kothari.
\newblock {\em Research methodology: Methods and techniques}.
\newblock New Age International, 2004.

\bibitem{tsang2014case}
Eric~WK Tsang.
\newblock Case studies and generalization in information systems research: A
  critical realist perspective.
\newblock {\em The Journal of Strategic Information Systems}, 23(2):174--186,
  2014.

\bibitem{benbasat1987case}
Izak Benbasat, David~K Goldstein, and Melissa Mead.
\newblock The case research strategy in studies of information systems.
\newblock {\em MIS quarterly}, pages 369--386, 1987.

\bibitem{xu2022short}
Teng~Andrea Xu and Jiahua Xu.
\newblock A short survey on business models of decentralized finance (defi)
  protocols.
\newblock {\em arXiv preprint arXiv:2202.07742}, 2022.

\bibitem{digiconomist_2022}
Digiconomist.
\newblock Bitcoin energy consumption index, Apr 2022.

\bibitem{kimmell2022}
Christopher Bendiksen;~Matthew Kimmell.
\newblock The bitcoin mining network, Jul 2022.

\bibitem{gallersdorfer2020energy}
Ulrich Gallersd{\"o}rfer, Lena Klaa{\ss}en, and Christian Stoll.
\newblock Energy consumption of cryptocurrencies beyond bitcoin.
\newblock {\em Joule}, 4(9):1843--1846, 2020.

\bibitem{stoll2019carbon}
Christian Stoll, Lena Klaa{\ss}en, and Ulrich Gallersd{\"o}rfer.
\newblock The carbon footprint of bitcoin.
\newblock {\em Joule}, 3(7):1647--1661, 2019.

\bibitem{de2020bitcoin}
Alex De~Vries.
\newblock Bitcoin’s energy consumption is underestimated: A market dynamics
  approach.
\newblock {\em Energy Research \& Social Science}, 70:101721, 2020.

\bibitem{lewenberg2015bitcoin}
Yoad Lewenberg, Yoram Bachrach, Yonatan Sompolinsky, Aviv Zohar, and Jeffrey~S
  Rosenschein.
\newblock Bitcoin mining pools: A cooperative game theoretic analysis.
\newblock In {\em Proceedings of the 2015 international conference on
  autonomous agents and multiagent systems}, pages 919--927. Citeseer, 2015.

\bibitem{onat2021bitcoin}
Nuri Onat, Rateb Jabbar, Murat Kucukvar, and Noora Fetais.
\newblock Bitcoin and global climate change: Emissions beyond borders.
\newblock 2021.

\bibitem{de2022revisiting}
Alex De~Vries, Ulrich Gallersd{\"o}rfer, Lena Klaa{\ss}en, and Christian Stoll.
\newblock Revisiting bitcoin’s carbon footprint.
\newblock {\em Joule}, 6(3):498--502, 2022.

\bibitem{khan2003five}
Khalid~S Khan, Regina Kunz, Jos Kleijnen, and Gerd Antes.
\newblock Five steps to conducting a systematic review.
\newblock {\em Journal of the royal society of medicine}, 96(3):118--121, 2003.

\bibitem{huebner2017we}
Gesche~M Huebner, Moira~L Nicolson, Michael~J Fell, Harry Kennard, Simon Elam,
  Clare Hanmer, Charlotte Johnson, and David Shipworth.
\newblock Are we heading towards a replicability crisis in energy efficiency
  research? a toolkit for improving the quality, transparency and replicability
  of energy efficiency impact evaluations.
\newblock UKERC, 2017.

\bibitem{tabachnick2007using}
Barbara~G Tabachnick, Linda~S Fidell, and Jodie~B Ullman.
\newblock {\em Using multivariate statistics}, volume~5.
\newblock pearson Boston, MA, 2007.

\bibitem{denis2015applied}
Daniel~J Denis.
\newblock {\em Applied univariate, bivariate, and multivariate statistics}.
\newblock John Wiley \& Sons, 2015.

\bibitem{jana2022taming}
Rabin~K Jana, Indranil Ghosh, and Martin~W Wallin.
\newblock Taming energy and electronic waste generation in bitcoin mining:
  Insights from facebook prophet and deep neural network.
\newblock {\em Technological Forecasting and Social Change}, 178:121584, 2022.

\bibitem{liu2021bitcoin}
Zhaoyan Liu, Binfa Chen, Zhen Li, Hui Li, Dengzheng Wang, Yang Lyu, Lu~Gao, and
  Bingxu Hou.
\newblock Bitcoin mining recognition based on community detection with
  electricity consumption data.
\newblock In {\em 2021 IEEE 5th Conference on Energy Internet and Energy System
  Integration (EI2)}, pages 3091--3096. IEEE, 2021.

\bibitem{gundaboina2021energy}
Lokesh Gundaboina, Sumit Badotra, and Sarvesh Tanwar.
\newblock Energy and resource consumption in cryptocurrency mining: A detailed
  comparison.
\newblock In {\em 2021 9th International Conference on Reliability, Infocom
  Technologies and Optimization (Trends and Future Directions)(ICRITO)}, pages
  1--5. IEEE, 2021.

\bibitem{platt2021energy}
Moritz Platt, Johannes Scdlmeir, Daniel Platt, Jiahua Xu, Paolo Tasca, Nikhil
  Vadgama, and Juan~Ignacio Iba{\~n}ez.
\newblock The energy footprint of blockchain consensus mechanisms beyond
  proof-of-work.
\newblock In {\em 2021 IEEE 21st International Conference on Software Quality,
  Reliability and Security Companion (QRS-C)}, pages 1135--1144. IEEE, 2021.

\bibitem{roma2020energy}
Crystal~Andrea Roma and M~Anwar Hasan.
\newblock Energy consumption analysis of xrp validator.
\newblock In {\em 2020 IEEE International Conference on Blockchain and
  Cryptocurrency (ICBC)}, pages 1--3. IEEE, 2020.

\bibitem{li2019energy}
Jingming Li, Nianping Li, Jinqing Peng, Haijiao Cui, and Zhibin Wu.
\newblock Energy consumption of cryptocurrency mining: A study of electricity
  consumption in mining cryptocurrencies.
\newblock {\em Energy}, 168:160--168, 2019.

\bibitem{coinmetrics_2020}
CoinMetrics.
\newblock Coin metrics' state of the network, Apr 2020.

\bibitem{stodden2010reproducible}
Victoria~C Stodden.
\newblock Reproducible research: Addressing the need for data and code sharing
  in computational science.
\newblock 2010.

\bibitem{zade2019bitcoin}
Michel Zade, Jonas Myklebost, Peter Tzscheutschler, and Ulrich Wagner.
\newblock Is bitcoin the only problem? a scenario model for the power demand of
  blockchains.
\newblock {\em Frontiers in Energy Research}, 7:21, 2019.

\bibitem{o2008information}
Patrick O'Beirne.
\newblock Information and data quality in spreadsheets.
\newblock {\em arXiv preprint arXiv:0809.3609}, 2008.

\end{thebibliography}
\appendix

\newpage
\section{Keywords Formulation: Snowballing\label{AppendixA}}
\setcounter{figure}{0} 
\begin{figure}[h]
    \centering
    \includegraphics[scale=0.29]{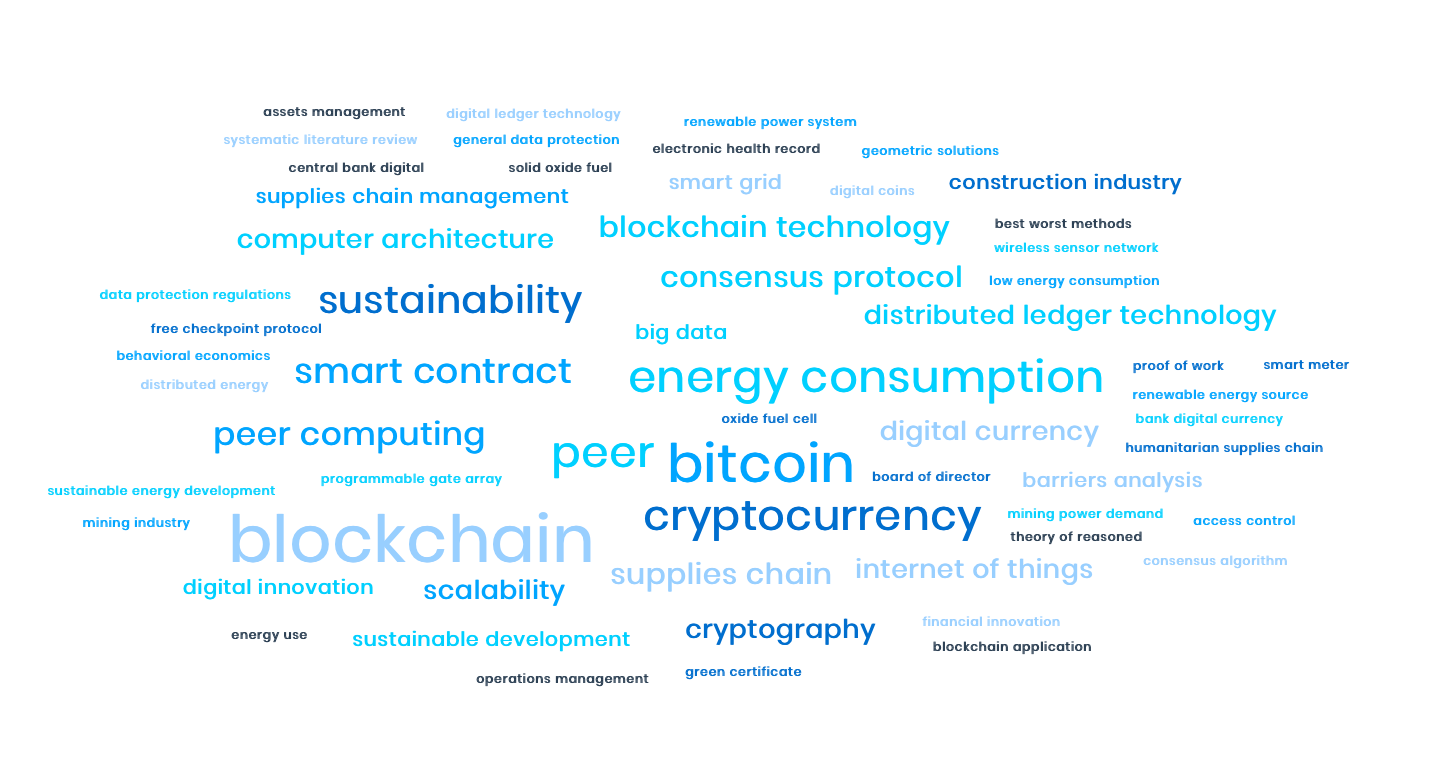}
    \caption{Word cloud of keywords obtained through forward and backward snowballing on Vranken (2017)}
    \label{fig:Keywords}
\end{figure}
\section{Search Queries \label{AppendixB}}
\begin{itemize}
\item \textit{\textbf{Google Scholar, IEEE Digital Library and Science Direct:} } 
\begin{lstlisting}
("Blockchain" OR "DLT" OR "bitcoin" OR "blockchain" OR "cryptocurrencies" OR "cryptocurrency" OR "digital currency" OR "distributed ledger" OR "peer-to-peer computing" OR "smart contract platform") AND ((("Energy" OR "electricity" OR "power" OR "power supply") AND ("Consumption" OR "expenditure" OR "use" OR "utilisation" OR "utilization")) OR ("Environment" OR "atmosphere" OR "carbon" OR "climate" OR "ecological" OR "emission" OR "environmental" OR "green" OR "footprint" OR "e-waste") OR ("Sustainability" OR "green design" OR "green technology" OR "sustainable"))
\end{lstlisting}

\item \textit{\textbf{ACM Digital Library:} }
\begin{lstlisting}
+("Blockchain" "DLT" "bitcoin" "blockchain" "cryptocurrencies" "cryptocurrency" "digital currency" "distributed ledger" "peer-to-peer computing" "smart contract platform") + ((("Energy" "electricity" "power" "power supply") + ("Consumption" "expenditure" "use" OR "utilisation" "utilization")) ("Environment" "atmosphere" "carbon" "climate" "ecological" "emission" "environmental" "green" "footprint" "e-waste") ("Sustainability" "green design" "green technology" "sustainable"))
\end{lstlisting}

\item \textit{\textbf{ISI Web of Science:} }
\begin{lstlisting}
(TS= ("Blockchain" OR "DLT" OR "bitcoin" OR "blockchain" OR "cryptocurrencies" OR "cryptocurrency" OR "digital currency" OR "distributed ledger" OR "peer-to-peer computing" OR "smart contract platform") AND ((("Energy" OR "electricity" OR "power" OR "power supply") AND ("Consumption" OR "expenditure" OR "use" OR "utilisation" OR "utilization")) OR ("Environment" OR "atmosphere" OR "carbon" OR "climate" OR "ecological" OR "emission" OR "environmental" OR "green" OR "footprint" OR "e-waste") OR ("Sustainability" OR "green design" OR "green technology" OR "sustainable")) AND LANGUAGE: (English) 
\end{lstlisting}
\item \textit{\textbf{Scopus:} }
\begin{lstlisting}
 (( TITLE-ABS-KEY("Blockchain" OR "DLT" OR "bitcoin" OR "blockchain" OR "cryptocurrencies" OR "cryptocurrency" OR "digital currency" OR "distributed ledger" OR "peer-to-peer computing" OR "smart contract platform") AND ((("Energy" OR "electricity" OR "power" OR "power supply") AND ("Consumption" OR "expenditure" OR "use" OR "utilisation" OR "utilization")) OR ("Environment" OR "atmosphere" OR "carbon" OR "climate" OR "ecological" OR "emission" OR "environmental" OR "green" OR "footprint" OR "e-waste") OR ("Sustainability" OR "green design" OR "green technology" OR "sustainable")) AND  PUBYEAR  >  2008  AND  ( LIMIT-TO ( LANGUAGE ,  "English" ) )
\end{lstlisting}

\item \textit{\textbf{Springer Link:}   } \begin{lstlisting}
("Blockchain" - "DLT" - "bitcoin" - "blockchain" - "cryptocurrencies" - "cryptocurrency" - "digital currency" - "distributed ledger" - "peer-to-peer computing" - "smart contract platform") & ((("Energy" - "electricity" - "power" - "power supply") & ("Consumption" - "expenditure" - "use" - "utilisation" - "utilization")) - ("Environment" - "atmosphere" - "carbon" - "climate" - "ecological" - "emission" - "environmental" - "green" - "footprint" - "e-waste") - ("Sustainability" - "green design" - "green technology" - "sustainable"))
\end{lstlisting}

\end{itemize}

\section{Reviewed Cryptocurrencies \label{AppendixC}}

\begin{table}[H]
\tiny

\end{table}
\newpage

\section{Shortlisted Studies \label{AppendixD}}

\footnotesize
\begin{longtable}{|r|p{8.5cm}|p{5.5cm}|}

\hline
No   & Study Title                                                                                                                                                         & Author/s                                                                                                                            \\ \hline
1    & Life Cycle Assessment of Bitcoin Mining                                                                                                                             & S. Köhler                                                                                                                           \\ \hline
2    & Block chain infrastructure and growth of global power consumption                                                                                                   & V. Denisova                                                                                                                         \\ \hline
3    & Energy Consumption Analysis of XRP Validator                                                                                                                        & C.A. Roma                                                                                                                           \\ \hline
4    & Blockchain and Ecological Impact: Between Reality and Accusation?                                                                                                   & M.B. Chenguel                                                                                                                       \\ \hline
5    & Bitcoin carbon footprint: Mining pools based estimate methodology                                                                                                   & K. Kononova                                                                                                                         \\
\hline
\end{longtable}

\end{document}